\documentclass[twocolumn,showpacs,prb,superscriptaddress,aps,floatfix]{revtex4-1}

\usepackage{rotating}
\usepackage{amsmath}
\usepackage{tabularx}
\usepackage{mathtools}
\usepackage{amssymb}
\usepackage{color}
\usepackage{graphicx}
\usepackage[active]{srcltx}
\usepackage[sort&compress]{natbib}
\usepackage{amssymb}
\usepackage[dvips]{epsfig}
\usepackage{float}

\renewcommand{\[}{\left[}
\renewcommand{\]}{\right]}
\renewcommand{\(}{\left(}
\renewcommand{\)}{\right)}

\newcommand{\be}{\begin{equation}}
\newcommand{\ee}{\end{equation}}
\newcommand{\bea}{\begin{eqnarray}}
\newcommand{\eea}{\end{eqnarray}}

\renewcommand{\v}[1]{{\bf #1}}

\def\nnpkkp {\begin{subarray}{l} n n'\\ \kk \kk' \end{subarray}}
\def\nmkp   {\begin{subarray}{l} n m\\ \kk \pp \end{subarray}}
\def\mnpk   {\begin{subarray}{l} m n\\ \pp \kk \end{subarray}}
\def\nnkk   {\begin{subarray}{l} n n\\ \kk \kk \end{subarray}}
\def\mnppkp {\begin{subarray}{l} m n'\\ \pp \kk' \end{subarray}}
\def\aimbpt {\textit{ai}--MBPT\,}
\def\limq	{\lim_{\qq\rightarrow{\bf 0}}}
\def\ai		{{\em ab--initio}}

\def\tt		{\boldsymbol{\tau}}

\def\pp		{{\bf p}}
\def\rr		{{\bf r}}

\def\whH		{\widehat{H}}
\def\whW		{\widehat{W}}

\def\FF		{{\bf F}}

\def\GG		{{\bf G}}

\def\RR		{{\bf R}}

\def\oRR        {\overline{\bf R}}

\def\dRu	{\Delta{R}}

\def\brank  { \la n\kk \(\RR\)|}
\def\ketmp  { | m\pp \(\RR\)\ra}
\def\PP		{{\bf P}}

\def\qq		{{\bf q}}
\def\kk		{{\bf k}}

\def\ga         {\alpha}
\def\gb         {\beta}

\def\gd         {\delta}

\def\gee        {\epsilon}
\def\gl         {\lambda}

\def\go         {\omega}

\def\gql        {\qq \gl}
\def\gqlp       {\qq' \gl'}

\def\gr         {\rho}

\def\bgS        {{\bf\Sigma}}

\def\gt         {\theta}

\def\ep         {\gee_{p}}

\def\capo       {\right.\\ \left.}

\def\la         {\langle}
\def\ra         {\rangle}

\renewcommand{\[}{\left[}
\renewcommand{\]}{\right]}
\renewcommand{\(}{\left(}
\renewcommand{\)}{\right)}

\def\nk         {n{\bf k}}

\def\npp        {n'{\bf p}}

\def\dk         {\frac{d\,{\bf k}}{\(2 \pi\)^3}}
\def\dk1        {\frac{d\,{\bf k}_1}{\(2 \pi\)^3}}

\def\dk         {\frac{d\,{\bf k}}{\(2 \pi\)^3}}

\def\Vscf       {V_{scf} }

\def\epi        {electron-phonon interaction }
\def\epip        {electron-phonon interaction.}

\def\ep         {electron-phonon }

\def\nk         {n{\bf k}}

\def\bverb      {\renewcommand{\baselinestretch}{1}\begin{verbatim}}
\def\everb  	{\end{verbatim}\renewcommand{\baselinestretch}{1.3}}

\begin{document}
\title{A many--body perturbation theory approach to the electron--phonon interaction with density--functional theory as a starting point}

\author{Andrea Marini}
\affiliation{Istituto di Struttura della Materia of the National Research Council, Via Salaria Km 29.3,
I-00016 Monterotondo Stazione, Italy} 
\affiliation{European Theoretical Spectroscopy Facilities (ETSF)} 

\author{S. Ponc\'e}
\affiliation{Universit\'e catholique de Louvain, Institute of Condensed Matter and Nanosciences,
NAPS Chemin des \'etoiles 8, bte L07.03.01, B-1348 Louvain-la-neuve, Belgium} 
\affiliation{European Theoretical Spectroscopy Facilities (ETSF)} 

\author{X. Gonze}
\affiliation{Universit\'e catholique de Louvain, Institute of Condensed Matter and Nanosciences,
NAPS Chemin des \'etoiles 8, bte L07.03.01, B-1348 Louvain-la-neuve, Belgium} 
\affiliation{European Theoretical Spectroscopy Facilities (ETSF)} 

\date{\today}
\begin{abstract}
The electron--phonon interaction plays a crucial role in many fields of physics and chemistry. Nevertheless, its actual
calculation by means of modern many--body perturbation theory is weakened by the use of model Hamiltonians that
are based on parameters difficult to extract from the experiments. 
Such shortcoming can be bypassed by using density--functional theory to evaluate the electron--phonon scattering amplitudes, phonon frequencies and electronic 
bare energies.
In this work, we discuss how a consistent many--body diagrammatic
expansion can be constructed on top of density--functional theory. 
In that context, the role played by screening and self--consistency when all the components of the electron--nucleus and nucleus--nucleus interactions are taken into account is paramount.  
A way to avoid overscreening is notably presented.
Finally, we derive cancellations rules as well as internal consistency constraints in order to draw a clear, sound and practical scheme to merge
many--body perturbation and density--functional theory.
\end{abstract}           


%71.38.-k: Electron-phonon interactions, electronic structure of solids
%71.15.Mb: Density-functional theory, condensed matter
%71.15.-m: Perturbation theory

\maketitle

In physics and chemistry the interaction between electronic and vibrational
degrees of freedom is at the origin of a multitude of phenomena. Focusing on
solid state physics, this coupling usually determines the electrical
 and thermal conductivity of metals as well as carrier lifetime in
doped semiconductors\cite{Ziman1960}. It also induces the transition to a
superconducting phase in many solids and nanostructures\cite{Mitsuhashi2010}.
The electron--nucleus coupling also plays a role in the renormalization of electronic
bands\cite{Allen1983}, carriers mobility in organic devices\cite{Gosar1966} and
dissociation at the donor/acceptor interface in organic
photovoltaic\cite{Tamura2008}.
This coupling can naturally interact with other couplings like the magnetic field leading, for example, to the spin-Seebeck effect\cite{Adachi2013}.

This effect is nowadays the subject of an intense research activity for its crucial role in new emerging fields of 
experimental and theoretical physics. The electron--nucleus coupling plays a crucial role in the relaxation and dissipation of 
photo--excited carriers in
pump and probe experiments~\cite{Bernardi1,Sangalli1}. Similarly, modern 
Angle--Resolved Photoemission Spectra\,(ARPES) experiments have recently disclosed the complex and temperature dependent structures appearing in the spectral functions of
several oxides\cite{Moser2013}. These structures quite remarkably resemble similar structures predicted to exist in conjugated polymers\cite{Cannuccia2011,Cannuccia2012}
pointing to a strong effect whose physical origin is still not completely clear.

From the theoretical point of view, the most up--to--date scheme to calculate and predict the ground-- and excited--state properties of 
a wide range of materials is based on the merging of
Density--Functional--Theory\,(DFT)\cite{R.M.Dreizler1990} with  Many--Body Perturbation
Theory\,(MBPT)\cite{Onida2002}. 

DFT is a broadly used \textit{ab-initio} ground--state theory,
that allows to calculate \textit{exactly} electronic density and total energy without adjustable parameter. 
The merging of DFT with perturbation theory gives the so-called 
Density-Functional Perturbation Theory\cite{Baroni1987,Gonze1997,baroni2001} (DFPT). The DFPT  is a powerful computational tool for the direct treatment of phonons.

However, the DFT computation of excited electronic states 
properties like the bandgap energies is a known problematic topic~\cite{R.M.Dreizler1990}.
As a result, MBPT
is nowadays the preferred alternative to DFT for that purpose. It
is based on the accurate treatment of correlation effects by means of the
Green's function formalism. MBPT is formally correct and leads to a close
agreement with experiment\cite{Schilfgaarde2006} but is extremely
computationally demanding. A natural way to solve this issue is to merge the
quick DFT calculation with the accurate MBPT one. The latter method
is often referred to as \textit{ab-initio} Many-Body Perturbation
Theory~\cite{Onida2002} (\textit{ai}--MBPT).  
In this method, DFT provides a suitable single--particle basis for the MBPT scheme.
This methods has been applied successfully to correct the well--known band--gap underestimation problem of
DFT\cite{Gruning2006,Niquet2004}.

Although the \aimbpt aims at calculating the excited state properties with an unprecedent precision, 
it is commonly applied by neglecting the effect of lattice vibrations.
Even today, most of the \textit{ai}--MBPT results are compared with finite-temperature experimental data\cite{Cardona1999a}.
Such comparison is not even well motivated at zero temperature as the lattice vibrations induce a zero--point motion effect
that can be sizeable~\cite{Cannuccia2011,Cannuccia2012,Gonze2011,Giustino2010},
e.g. on the order of 0.4--0.6 eV for the direct and indirect band gaps of diamond~\cite{Ponce2014,Antonius2014}. 
This represent a clear motivation to develop a coherent \ai\, theory in which the electron--phonon interaction is rigorously included on top
of \textit{ai}--MBPT.

The need for such theory is exemplified by the very fragmented historical development of the \ai\, approach to the temperature dependence
of the electronic structure due to the electron--phonon interaction.
From the fifties to the late eighties, a coherent \ai\, framework was still not devised and the electron--phonon interaction 
was initially investigated and computed in a semi-empirical context
by Fan\cite{Fan1950,Fan1951}. His theory had no adjustable parameters and was based on the first-order perturbed 
Hamiltonian. During the same period, Anton\v{c}\'{i}k\cite{Antoncik1955}, followed by others\cite{Keffer1968,Walter1970,Kasowski1973}, developed empirical Debye-Waller\,(DW) corrections to the nuclear potential. 
Only in 1976, Allen and Heine\cite{Allen1976} rigorously unified the Fan and DW corrections in a common framework. Their approach, combined with the use of a semi-empirical 
model, allows for a re-writing of the problem in terms of first-order derivatives of an effective potential only. Calculations of the
electron--phonon renormalization effects were then led by Cardona and 
coworkers\cite{Allen1981,Allen1983,Lautenschlager1985,Zollner1992}, including Allen.
The resulting approach is now called the Allen-Heine-Cardona\,(AHC) theory.

In 1989, the first \ai\, calculation of the temperature dependence of the gap was attempted, by King-Smith \textit{et~al}\cite{King-Smith1989}, based on DFPT. 
Starting from there, several
first--principle calculations have been done, relying mainly on three types of formalisms: (i) time averaging of bandgap obtained using first principles molecular dynamics 
simulations\cite{Franceschetti2007,Kamisaka2008,Ibrahim2008,Ramirez2006,Ramirez2008}; (ii) frozen phonons\,(FP) calculations~\cite{Monserrat2013,Antonius2014,Capaz2005,Patrick2013,Han2013,SP_2014} and
(iii) the AHC approach implemented in a full \ai\, framework by using DFT and DFPT as a reference system~\cite{Giustino2010,Ponce2014,Antonius2014,SP_2014,Kawai2014}.

All these approaches are based on an adiabatic and static treatment of the electron--phonon interaction. This limitation was overcome by using
the dynamical version of MBPT by Marini~\textit{et~al}\cite{Marini2008,Cannuccia2011,Cannuccia2012} who focused on retardation effects.

Since then, there have been an increasing number of studies in which the electron--phonon interaction is fully included 
in the computation of the electronic structure, well beyond DFT. Still, several basic questions remain. In particular,
the use of an electron--phonon interaction whose strength is computed from DFT, in a formalism that goes beyond DFT, e.g. the \aimbpt approach, leads to 
several ambiguities due to the simultaneous inclusion of different levels of correlation at the MBPT and DFT/DFPT level. 

Indeed, the DFPT electron--phonon interaction is naturally screened as it is computed from the derivative of the \textit{self--consistent} Kohn-Sham potential with respect to atomic displacements. 
This screening is taken as it is in the MBPT part of the \aimbpt scheme, although it is well known that the diagrammatic technique also predicts the screening of
the electron--phonon interaction consistently with the kind of correlation included in the self--energy~\cite{mattuck,ALEXANDERL.FETTER1971}.
It has in fact been shown that the size of the zero-point motion renormalization is significantly larger in the MBPT than in the DFPT approaches\cite{Antonius2014}.

Another important issue of the \aimbpt approach  is the lack of a diagrammatic interpretation of the screening of the Debye--Waller term. This screening arises quite naturally
in the DFPT~\cite{SP_2014} and AHC approaches~\cite{Allen1976}. It is easy to show that it comes from the DFT self--consistent screened ionic potential. Instead, in the pure MBPT treatment
of electrons and nuclei, this diagram is un--screened.
The Debye--Waller diagram is however usually taken as screened without justification in most practical application because of the DFPT basis.

In addition, a non--rigid nuclei correction to the Debye--Waller contribution~\cite{SP_2014} is predicted to exist within the DFPT approach. However, this term is notably absent from the standard derivation of the electron--phonon theory based on the MBPT. 

The last issue is even more fundamental. Most of the electron--phonon interaction treatments that appears in textbooks, see e.g. Refs.~[\onlinecite{mahan,ALEXANDERL.FETTER1971,mattuck}], are based on the study of the homogeneous electronic gas~(jellium). At variance with any realistic
material, the jellium model is based on a drastic approximation: 
the ions are replaced by a jelly of positive charge, in contrast with realistic materials where the nuclei and their mutual interaction must be taken explicitly into account. This is correctly done in DFT and DFPT but not in the MBPT approach derived from the jellium model. 

This paper aims at answering all these questions by devising a coherent, formal and accurate approach to merge the MBPT scheme with DFT and DFPT.
We  present a consistent electron--phonon interaction theory based on the MBPT formalism, insisting specifically on the connection between the MBPT and AHC approaches. This work is inspired by the seminal works of 
Allen\cite{Allen1978a} and van Leeuwen\cite{Leeuwen2004a}, going further by including the full description of the atomic potential into account.

The merging procedure will lead to the natural definition of a series of practical rules and advices about how to perform electron--phonon calculations on top of DFT without 
double counting problems. These series of {\em rules} are well justified within the  \aimbpt scheme that, in its practical form used in material science calculations,
can be seen as a collection of prescriptions only partially based on a solid theoretical ground and rather inspired by the succesfull comparison with the experiments
of several different materials. This {\em a posteriori} validation represents and important part of the \aimbpt approach.

The structure of the paper is as follows. Section~\ref{sec:the_hamiltonian} presents the total Hamiltonian and introduces the notation. In section~\ref{sec:the_problem}, we draw a parallel 
between the electron--electron and the electron--phonon self--energies to show what is the source of the problems that arise in the
merging of MBPT with DFT and DFPT. Section~\ref{sec:the_ip_Ho} properly defines the reference Hamiltonian to be used as a zero--th order in the many--body expansion.
In Section~\ref{sec:the_interaction}, the different interaction terms are described, including the contributions from the nuclei--nuclei interaction. In
Section~\ref{sec:the_expansion}, we perform the formal diagrammatic summations at different level of approximations: Hartree, Hartree--Fock and $GW$. 
We use the different levels of correlation of these self--energies to discuss
the different role played by self--consistency diagrams and how the screening of the interaction terms arises. 
At the same time we derive cancellation rules that highlight the crucial role
played by the nuclei--nuclei interaction. 

Finally, Section~\ref{sec:the_dfpt} reviews the DFPT approach to the electron--phonon coupling in order, in Section~\ref{sec:merge}, to compare the different properties of the DFT and of the Many--Body approach.
We provide, in a practical and schematic way, a series of formal properties of the Many--Body expansion performed on top of the DFT reference Hamiltonian. We discuss, from a diagrammatic perspective, the 
physical origin of the Debye--Waller terms beyond the screened rigid--ion contribution  (Section~\ref{sec:nddw}) and a practical approach to calculate iteratively the n--th order derivatives of the DFT self--consistent potential
(Section~\ref{sec:nth_order} and Appendix~\ref{appA}). Atomic (Hartree) units are used throughout the article.

%%%%%%%%%%%%%%%%%%%%%%%%%%%%%%%%%%%%%%%%%%%%%%%%%%%%%%
\section{The total Hamiltonian}
\label{sec:the_hamiltonian}
%%%%%%%%%%%%%%%%%%%%%%%%%%%%%%%%%%%%%%%%%%%%%%%%%%%%%%
We start from the generic form of the total Hamiltonian of the system, that we divide in its electronic $\widehat{H}_e$,
nuclear $\widehat{H}_n\(\RR\)$ and electron--nucleus\,(e--n) $\whW_{e-n}\(\RR\)$ contributions,
\begin{align}
\widehat{H}\(\RR\)= \widehat{H}_e+\widehat{H}_n\(\RR\)+\whW_{e-n}\(\RR\),
\label{eq:1.0}
\end{align}
where $\RR$ is a generic notation that represents a dependence on the positions of the nuclei.

The electronic and nuclear parts are divided in a kinetic $\widehat{T}$ and interaction part
$\whW$:
\begin{gather}
 \widehat{H}_e =\widehat{T}_e+\whW_{e-e},
\label{eq:1.0a}\\
 \widehat{H}_n\(\RR\) =\widehat{T}_n+\whW_{n-n}\(\RR\).
\label{eq:1.0b}
\end{gather}
Note that the nuclear kinetic energy depends on the nuclear momenta, and not on the actual positions of the nuclei.
In the above definitions, the operators are bare (un--dressed). The analysis of the dressing of $\whW_{e-n}$ that arises as a consequence of the electronic correlations is one of the key objective of this work.
Indeed, in the Many--Body (MB) approach, this dressing appears in the perturbative expansion in the form of electron--hole pair excitations and therefore, cannot be introduced \textit{a priori}
in the definition of the Hamiltonian.

The explicit expression for the bare (e--n) interaction term is
\begin{multline}
\whW_{e-n}\(\RR\)= \\
-\sum_{ls,i}\frac{Z_s}{|\widehat{\rr}_i-\widehat{\RR}_{ls}|}=-\sum_{ls,i}Z_s v\(\widehat{\rr}_i-\widehat{\RR}_{ls}\),
\label{eq:1.1a}
\end{multline}
where $\widehat{\RR}_{ls}$ is the nuclear position operator for the nucleus $s$
inside the cell $l$ (the cell is located at position $\RR_l$), $Z_s$ is the corresponding charge, $\widehat{\rr}_i$ is the electronic position operator of the electron $i$ and
$v\(\rr-\rr'\)=|\rr-\rr'|^{-1}$ is the
bare Coulomb potential. Similarly,
\begin{gather}
\whW_{n-n}\(\RR\)=\frac{1}{2}\sum_{ls,l's'}\nolimits'Z_s Z_{s'}v\(\widehat{\RR}_{ls}-\widehat{\RR}_{l's'}\),
\label{eq:1.1b}\\
\whW_{e-e}=\frac{1}{2}\sum_{i j}\nolimits'v\(\widehat{\rr}_i-\widehat{\rr}_j\),
\label{eq:1.1c}
\end{gather}
with $\sum_{ij}\nolimits'=\sum_{i\neq j}$. 

We now use the notation $\overline{O\(\RR\)}$, or equivalently $O\(\overline{\RR}\)$, to indicate a quantity or an operator that is evaluated 
with the nuclei frozen in their equilibrium crystallographic positions ($\overline{\RR}$). We expand the Hamiltonian as a Taylor series up to second order in the 
nuclear displacements,
\begin{multline}
\widehat{H}\(\RR\) \approx \widehat{H}\(\overline{\RR}\) 
+\sum_{ls\alpha}\overline{\partial_{R_{ls\alpha}} H\(\RR\)} \Delta \widehat{R}_{ls\alpha} \\
+\frac{1}{2}\sum_{ls\alpha,l's'\beta}\overline{\partial^2_{R_{ls\alpha}R_{l's'\beta}} H\(\RR\)}
\Delta \widehat{R}_{ls\alpha}\Delta \widehat{R}_{l's'\beta},
\label{eq:1.4}
\end{multline}
where  $\alpha$ and $\beta$ are Cartesian coordinates and 
\begin{align}
\Delta \widehat{R}_{ls\alpha}\equiv \( \widehat{R}_{ls\alpha} -\overline{R}_{ls\alpha}\hat{\mathbf{1}}\).
\end{align}

The equilibrium crystallographic positions $\overline{\RR}$ are defined, in the present context, as the 
positions minimizing the expectation energy of the Born-Oppenheimer Hamiltonian (with fully correlated electrons), i.e. all the contributions
to the total Hamiltonian, except the nuclear kinetic energy,
\begin{align}
\widehat{H}_{BO}\(\RR\)=\widehat{H}_e+\widehat{W}_{n-n}\(\RR\) +\widehat{W}_{e-n}\(\RR\).
\end{align}
 Those positions are equivalently defined by the condition that
the expectation of the Born-Oppenheimer force $\FF_{\RR_{ls}}$ acting on the nucleus located at position $\RR_{ls}$ is zero
\begin{align}
\FF_{\RR_{ls}} \equiv -\partial_{\RR_{ls}} \left.\la  \widehat{H}_{BO}\(\RR\)  \ra\right|_{\RR_{ls}=\overline{\RR}_{ls}}  
= 0\quad\forall \{l,s\}.
\label{eq:1.4s}
\end{align}
The average in Eq.~\eqref{eq:1.4s} is done on the exact electronic ground state of the Born-Oppenheimer Hamiltonian. Still, the present theory
will go beyond the Born-Oppenheimer approximation by considering fluctuations around the equilibrium positions.

%%%%%%%%%%%%%%%%%%%%%%%%%%%%%%%%%%%%%%%%%%%%%%%%%%%%%%%%%%%%%%%%%%%%%%%%%%%%%%%%%%%%%%%%%%%%%%%%%%%%%%%%%%%%%
\section{The problem}
\label{sec:the_problem}
%%%%%%%%%%%%%%%%%%%%%%%%%%%%%%%%%%%%%%%%%%%%%%%%%%%%%%%%%%%%%%%%%%%%%%%%%%%%%%%%%%%%%%%%%%%%%%%%%%%%%%%%%%%%%
The problem we aim at solving is how to treat the effect of the two last terms in the right--hand side of Eq.~\eqref{eq:1.4} and how to do it
by merging the MB approach, well-established for the treatment of $\widehat{H}_{BO}\(\overline{\RR}\)$, with a DFT description of the reference electronic and nuclear systems.

When $\widehat{H}\(\RR\)\approx\widehat{H}_{BO}\(\overline{\RR}\)$ the Hamiltonian represents indeed
a purely electronic problem, for which the MB approach is well-established in the 
literature\cite{Onida2002,mahan,Hedin19701,ALEXANDERL.FETTER1971}. It relies on the definition of an electronic self--energy
$\Sigma\(\rr,\rr';\go\)$ that is a complex and non--local function in frequency and space. $\Sigma$ can be approximated by following different strategies available 
in the literature (like the well--known $GW$ approximation~\cite{gunnarson1998}). 
For periodic solids, once the self--energy is known, the calculation of the correction to an energy level $|n\kk\ra$ can be obtained by solving the corresponding
Dyson equation ($n$ is a band index and $\kk$ the corresponding wavevector). 

Usually, the MB methodology starts from an independent--particle\,(IP) electronic Hamiltonian  that includes only the kinetic 
electronic operator and the electron-nucleus operator,
\begin{align}
\widehat{H}_{IP}\(\overline{\RR}\)\equiv \widehat{T}_e+\widehat{W}_{e-n}\(\overline{\RR}\).
\label{eq:1.5a}
\end{align}
The analysis of the correlated electronic Hamiltonian, 
\begin{align}
\widehat{H}_{corr}\(\overline{\RR}\) =  \widehat{H}_{IP}\(\overline{\RR}\) + \widehat{W}_{e-e},
\label{eq:1.5b}
\end{align}
is addressed through the diagrammatic expansion.
A simple approximation to the solution of the Dyson Equation that fully captures the role played by correlation effects
is the 
on--the--mass--shell approximation
where:
\begin{align}
\gee_{n\kk}\approx \gee^{\(0\)}_{n\kk}+\prescript{}{0}{\la n\kk|}\Sigma\(\rr,\rr'; \gee^{\(0\)}_{n\kk}\)| n\kk\ra_0,
\label{eq:1.5}
\end{align}
where $|n\kk\ra_0$ and $\gee^{\(0\)}_{n\kk}$ are the n--th single--particle eigenstate
and eigenenergy of the independent--particle\,(IP) Hamiltonian $\widehat{H}_{IP}\(\overline{\RR}\)$.

In the present context, where we must consider different configurations of nuclei, and determine 
also the equilibrium geometry through Eq.~\eqref{eq:1.4s}, the initial correlation present in the reference system for the diagrammatic expansion must be carefully analyzed.
Adding the nucleus-nucleus energy to $\widehat{H}_{IP}\(\RR\)$ gives
\begin{align}
\widehat{H}_0\(\RR\)\equiv \widehat{T}_e+ \widehat{W}_{e-n}\(\RR\) +\widehat{W}_{n-n}\(\RR\),
\label{eq:1.5cc}
\end{align}
namely, the Born-Oppenheimer Hamiltonian without electron--electron interaction operator  $\widehat{W}_{e-e}$.
This initial Hamiltonian $\widehat{H}_0$ do not include electron--electron correlations, and can be used as the starting point of a MB approach to the electronic problem.
However, using this Hamiltonian instead of the true Born-Oppenheimer Hamiltonian
means that no electronic correlation energy contribution
appears in the total energy and in the definition of the equilibrium nuclear positions through Eq.~\eqref{eq:1.4s}. This would lead 
to a completely irrealistic description of the starting nuclear geometry and vibrational frequencies. Indeed, e.g. the latter could be imaginary,
and this would lead to unusual technical problems with the canonical transformation from the displacement operator to the
phonon creation and annihilation operators.
Thus, the extension of the electronic--only MB approach to the case where nuclear displacements are considered cannot use such a starting point.

As DFT provides a treatment of energy and forces that include the electron--electron interaction, it yields a better starting point than Eq.~\eqref{eq:1.5cc}. 
DFT is an exact mean--field theory in the sense that
all electronic correlation effects are embodied in a mean--field exchange--correlation\,(xc) potential $V_{xc}\[\gr\]\(\widehat{\rr}\)$ which replaces the full electron--electron interaction operator $\whW_{e-e}$, and depends on the ground-state density $\gr$. The bracket $[\,]$ in $\widehat{V}_{xc}$  denotes a functional dependence.  
 
 By adding to $V_{xc}\[\gr\]\(\widehat{\rr}\)$ the Hartree potential, 
$V_{H}\[\gr\]\(\widehat{\rr}\)$, we get the total DFT potential:
\begin{align}
V_{Hxc}\[\gr\]\(\widehat{\rr}\)=\sum_i V_{H}\[\gr\]\(\widehat{\rr}_i\)+V_{xc}\[\gr\]\(\widehat{\rr}_i\),
\label{eq:1.6}
\end{align}
with 
\begin{align}
V_{H}\[\gr\]\(\widehat{\rr}\)=\int\,d\rr' v\(\widehat{\rr}-\rr'\) \gr\(\rr'\).
\label{eq:1.6p}
\end{align}
DFT is exact in the sense that the corresponding Kohn--Sham\,(KS) Hamiltonian
\begin{align}
\widehat{H}_{KS}\(\RR\)= \widehat{T}_e +\widehat{V}_{Hxc}\[\gr\] +\widehat{W}_{e-n}\(\RR\),
\label{eq:1.7}
\end{align}
provides, when the nuclear positions are given, a set of electronic eigenvectors whose corresponding density is the exact ground state density 
of $\widehat{H}_{BO}\(\RR\)$\,(Hohenberg--Kohn theorem\cite{R.M.Dreizler1990}). 

The Hohenberg--Kohn theorem also states that $\widehat{V}_{Hxc}[\rho]$ and the ground--state energy are functional of the exact 
electronic density $\rho$. 
It follows that, once the correct exchange-correlation functional is used, 
DFT gives the exact equilibrium nuclear positions through Eq.~\eqref{eq:1.4s}.

In practice, an exact expression for $\widehat{V}_{Hxc}$ is not known and 
several approximations for it have been proposed in the literature~\cite{R.M.Dreizler1990}. 
In any case, even the simple local--density approximation\,(LDA)~\cite{Ceperley1980,Perdew1981}, provides quite reasonable structural properties. Thus DFT represents a concrete and accurate 
reference Hamiltonian to be used as zero--th order for a diagrammatic expansion that will allow vibrational degrees of freedom to be included. 
Formally, at the equilibrium geometry, one decomposes the correlated Hamiltonian as 
\begin{align}
\widehat{H}_{corr}\(\overline{\RR}\)= \widehat{H}_{KS}\(\overline{\RR}\)+\whW_{e-e} - \widehat{V}_{Hxc}[\overline{\gr}].
\label{eq:1.8}
\end{align}
At this point, the perturbative expansion is performed in terms of $\whW_{e-e}-\widehat{V}_{Hxc}$ instead of $\whW_{e-e}$. This is the theoretical basis of the
standard \aimbpt\, scheme\cite{Onida2002}.

If DFT is used as a reference non--interacting system Eq.~\eqref{eq:1.5} does not hold anymore. Its extension can be shown to be 
\begin{multline}
\gee_{n\kk}\approx \gee^{KS}_{n\kk}
+ \prescript{}{KS}{\la n\kk|}\[\Sigma_{xc}\(\rr,\rr';\gee^{KS}_{n\kk}\)\right.\\
\left.-V_{xc}\[\overline{\gr}\]\(\widehat{\rr}\)\]| n\kk\ra_{KS},
\label{eq:1.9}
\end{multline}
where $|n\kk\ra_{KS}$ is the n--th single--particle eigenstate of $ \widehat{H}_{KS}$ with energy $ \gee^{KS}_{n\kk}$.  
Note that in Eq.~\eqref{eq:1.9} only the $\Sigma_{xc}$ and $V_{xc}$ terms appears as the Hartree terms in $\Sigma$ and $V_{Hxc}$ cancels out.

Eq.~\eqref{eq:1.9} reveals the simplicity of the \aimbpt\, scheme. The accuracy
and universality of DFT avoids the use of \textit{ad--hoc} parameters and the prize to pay (at least in the electronic case) is to simply subtract from the self--energy
the xc potential. This simplicity represents one of the key reasons for the wide--spread use of \aimbpt.

At this point, one would be tempted to follow the same strategy 
in the electron--phonon case by adding to Eq.~\eqref{eq:1.8}, the nuclear Hamiltonian, $\whH_{n}$, and the contributions from $\whW_{e-n}$  that are linear and quadratic in the atomic displacements. 
This is, however, formally not correct. Indeed, when the nuclei are allowed to be displaced from their equilibrium configuration, the DFT (or more directly, DFPT) 
will be expanded in a Taylor series, 
\begin{align}
\widehat{H}_{KS}\(\RR\)\approx\widehat{H}_{KS}\(\overline{\RR}\)+ \Delta{\widehat{H}_{KS}\(\RR\)},
\label{eq:1.10}
\end{align}
with, however,
\begin{align}
 \Delta{\widehat{H}_{KS}\(\RR\)} \neq  \widehat{H}\(\RR\) - \widehat{H}\(\overline{\RR}\).
\label{eq:1.10a}
\end{align}
This is due to the fact that, in $ \Delta{\widehat{H}_{KS}\(\RR\)}$, the \ep interaction terms are statically screened by the electronic dielectric function
while in the Taylor expansion Eq.~\eqref{eq:1.4} they are bare, unscreened. In other words, 
the ground state density $\rho$ present in Eq.~\eqref{eq:1.7} actually depends implicitly on the nuclear coordinates.
In practice, this means that the effect of $\widehat{V}_{Hxc}\[\overline{\gr}\]$ does not appear only as an additive term in the Dyson equation but it
screens the interaction potentials $\whW_{e-n}$ and $\whW_{n-n}$. 

An additional problem, partially connected to the potential double counting of correlation when DFT is used as the reference  Hamiltonian, is due to the fact that
most of the \ep\,theory has been devised in the jellium model where the nuclei appear only as static and frozen positive charges. 
As a consequence, strong approximations on the perturbative expansion are used in textbooks.
This is inconsistent with the microscopic description of the nuclear lattice and indeed represents the most critical shortcoming of the commonly applied
approaches. In all the aforementioned applications of the \aimbpt\, schemes (AHC and beyond) $\whW_{n-n}$ is neglected and $\whW_{e-n}$ is screened by hand directly in the initial Hamiltonian.

From these simple arguments we can argue that, although \textit{ai}--MBPT is a well-established scheme\cite{Onida2002}, its extension to the \ep problem is still far from being formally defined.

We would like to apply the MBPT technique to the perturbative expansion of Eq.~\eqref{eq:1.4} where the non--interacting Kohn--Sham Hamiltonian and its derivatives as calculated by DFT and DFPT are used.

We propose to apply the standard diagrammatic MBPT on the total bare Hamiltonian, given by Eqs.~\eqref{eq:1.4},
 explicitly taking into account all interaction terms. We will then examine the properties of the $\Delta{\widehat{H}_{KS}}$ operator in order to
draw a clear and formal comparison between \textit{ai}--MBPT and DFPT. 

In this way, we aim at creating a consistent framework where 
the role played by screening and self--consistency is clearly evidenced even when all the components of the electron--nucleus and nucleus--nucleus interactions are taken into account.

%%%%%%%%%%%%%%%%%%%%%%%%%%%%%%%%%%%%%%%%%%%%%%%%%%%%%%%%%%%%%%%%%%%%%%%
\section{The reference, independent--particle Hamiltonian}
\label{sec:the_ip_Ho}
%%%%%%%%%%%%%%%%%%%%%%%%%%%%%%%%%%%%%%%%%%%%%%%%%%%%%%%%%%%%%%%%%%%%%%%
As it emerges from the discussion of the previous section, the choice of the non--interacting Hamiltonian to be used as a reference for the perturbative expansion represents the
connection with DFT and thus provides the \ai\, basis for the entire theoretical derivation. This is particularly important for the definition of the phonon modes.
Therefore, we start by introducing a splitting of the total Hamiltonian in an independent particle part
(for independent electrons as well as independent phonons) plus interaction terms,
\begin{align}
\widehat{H}\(\RR\)=\widehat{H}_0\(\RR\)+ \Delta\widehat{H}\(\RR\).
\label{eq:ip.1}
\end{align}
Eq.\eqref{eq:ip.1} is more suited than Eq.~\eqref{eq:1.4} to the MB treatment.
The reference independent--particle Hamiltonian is 
\begin{multline}
\widehat{H}_0\(\RR\)=\widehat{T}_e+\widehat{T}_n+\whW_{e-n}\(\overline{\RR}\)\\
+\widehat{W}_{n-n}(\overline{\RR})+\Delta\whW^{ref}_{n-n}\(\RR\) ,
\label{eq:ip.2}
\end{multline}
where $\whW_{e-n}$ and $\widehat{W}_{n-n}$ are evaluated
at the equilibrium geometry.
We have introduced a reference nucleus--nucleus interaction $\Delta\whW^{ref}_{n-n}$, a second--order contribution
within the Taylor expansion in the nuclear displacements, that provides the reference phonon modes to be used in the diagrammatic expansion.
This $\Delta\whW^{ref}_{n-n}$ can be defined from DFPT
\begin{multline}
\Delta\whW^{ref}_{n-n}\(\RR\)=\\
\frac{1}{2}\sum_{ls\alpha,l's'\beta} \overline{\partial_{R_{ls\alpha}R_{l's'\beta}}^2 E^{BO}\(\RR\)}  \Delta \widehat{R}_{ls\alpha} \Delta \widehat{R}_{l's'\beta},
\label{eq:ip.4}
\end{multline}
with $E^{BO}$ the Born--Oppenheimer total energy of the system calculated within DFT. By construction, the phonon frequencies
and eigenvectors will be equal to those of the Born-Oppenheimer approximation based on the correlated electronic Hamiltonian.
Therefore, Eq.~\eqref{eq:ip.2} defines an independent--particle Hamiltonian beyond the equilibrium geometry.
The remaining interaction part, up to second order in nuclear displacements, is
\begin{multline}\label{delta_Hamil}
\Delta\widehat{H}\(\RR\)= \whW_{e-e}+ \Delta\whW_{e-n}\(\RR\)\\
+\Delta\whW_{n-n}\(\RR\)-\Delta\whW^{ref}_{n-n}\(\RR\).
\end{multline}
where
\begin{align}\label{delta_Wen}
\Delta\whW_{e-n}\(\RR\)=\whW_{e-n}\(\RR\)-\whW_{e-n}\(\overline{\RR}\),
\end{align}
and
\begin{align}\label{delta_Wen2}
\Delta\whW_{n-n}\(\RR\)=\whW_{n-n}\(\RR\)-\whW_{n-n}\(\overline{\RR}\).
\end{align}

At this point we can introduce the eigenstates of the nuclear harmonic oscillations of $\widehat{H}_0$, written in terms of
the canonical transformation
\begin{multline}
\Delta\hat{R}_{ls\alpha}=
\sum_{\qq \gl} \(2 N M_s \go_{\qq \gl}\)^{-1/2} \eta_{\ga}\(\qq\gl|s\)\\
\times e^{i \qq\cdot\oRR_{ls}}
\(\hat{b}^{\dagger}_{-\qq \gl}+\hat{b}_{\qq \gl}\),
\label{eq:2.1}
\end{multline}
where $\(\qq,\gl\)$ is a generic DFPT phonon mode with momentum $\qq$, energy branch $\gl$
and energy $\go_{\qq \gl}$.
$N$ is the number of $\qq$--points in the whole Brillouin Zone\,(BZ). We assume the $\qq$--point grid to be uniform so that
we have also $N$ $\kk$--points for the single particle representation.
$M_s$ is the nuclear mass, 
$\eta_{\alpha}\(\qq \gl | s\) $ is the polarization vector of the atom $s$ in the unit cell $l$ in the Cartesian direction $\alpha$,
while  $\hat{b}_{\qq \gl}$ and $\hat{b}^{\dagger}_{\qq \gl}$, respectively, are the annihilation and creation operators
of the phonon mode $\(\qq,\gl\)$, respectively.

We now introduce a second quantization formulation for the electrons. 
If $\phi_{n\kk}\(\rr\)$ is the $\widehat{H}_0$ electronic eigen--function, we introduce the field operator
\begin{align}
\widehat{\psi}\(\rr\)=\frac{1}{\sqrt{N}}\sum_{\kk} \widehat{\psi}_{\kk}\(\rr\),
\label{eq:2.1a}
\end{align}
with
\begin{align}
\widehat{\psi}_{\kk}\(\rr\)=\sum_n \phi_{n\kk}\(\rr\) \widehat{c}_{n\kk},
\label{eq:2.4a}
\end{align}
with $ \widehat{c}_{n\kk}$ the annihilation operator of an electron.
By using field operators, the independent-particle Hamiltonian can be written in a second quantization form:
\begin{align}
\widehat{H}_0\(\RR\)=
\sum_{\nk} \gee_{\nk} \hat{c}^{\dagger}_{\nk} \hat{c}_{\nk}+\sum_{\qq\gl}\go_{\qq\gl}\(\hat{b}^{\dagger}_{\qq \gl}\hat{b}_{\qq \gl}+\frac{1}{2}\), 
\label{eq:2.3}
\end{align}
where the first term corresponds to $\widehat{T}_e+\whW_{e-n}\(\overline{\RR}\)$ and the second one to 
$\widehat{T}_n\(\RR\)+\Delta\whW^{ref}_{n-n}\(\RR\)$. The Born-Oppenheimer energy of the ground state at equilibrium geometry 
has been redefined to be the zero of energy 
(hence e.g. $\widehat{W}_{n-n}(\overline{\RR})$ disappears from this expression).

The introduction of a reference nucleus--nucleus potential $\Delta\whW^{ref}_{n-n}\(\RR\)$ in $\widehat{H}_0\(\RR\)$ was the crucial step to be able to map it with its second quantization form. 
Indeed it is well known that phonon dynamics is actually decoupled from the electronic one, as shown in many references\cite{Leeuwen2004a,mattuck}.
Nonetheless, the phonon dynamics can describe accurately the vibrational properties of a real system only if it feels the electronic screening. Such screening is accounted for in DFPT by the reference potential $\Delta\whW^{ref}_{n-n}\(\RR\)$ operator. 

%%%%%%%%%%%%%%%%%%%%%%%%%%%%%%%%%%%%%%%%%%%%%%%%%%%%%%%%%%%%%%%%%%%%%%%
\section{The electron--phonon interaction terms}
\label{sec:the_interaction}
%%%%%%%%%%%%%%%%%%%%%%%%%%%%%%%%%%%%%%%%%%%%%%%%%%%%%%%%%%%%%%%%%%%%%%%
Thanks to the definition of the reference part of the Hamiltonian, Eq.~\eqref{eq:ip.2},  we have that the final 
splitting in independent--particle and interaction terms easily follows from Eq.~\eqref{eq:1.4}. The first two orders of the Taylor expansion of $\widehat{H}$ are:
\begin{align}
\Delta{\widehat{H}\(\RR\)}=\whW_{e-e}+\Delta\widehat{H}^{\(1\)}\(\RR\)+\Delta\widehat{H}^{\(2\)}\(\RR\),
\label{eq:2.3a}
\end{align}
with 
\begin{multline}
\Delta\widehat{H}^{\(1\)}\(\RR\)\equiv \\
\sum_{ls\ga}\nolimits\overline{\partial_{R_{ls\ga}} \[ W_{e-n}\(\RR\) + W_{n-n}\(\RR\) \]} \Delta \widehat{R}_{ls\ga},
\label{eq:2.4b}
\end{multline}
and
\begin{multline}
\Delta\widehat{H}^{\(2\)}\(\RR\)\equiv \\
\frac{1}{2}\sum_{ls\ga,l's'\gb}\overline{\partial^2_{R_{ls\ga}R_{l's'\gb}}
\[ W_{e-n}\(\RR\)+ W_{n-n}\(\RR\) \]} \\
 \times \Delta \widehat{R}_{ls\ga} \Delta \widehat{R}_{l's'\gb} -\Delta\whW^{ref}_{n-n}\(\RR\).
\label{eq:2.4c}
\end{multline}
By using Eq.~\eqref{eq:2.1} we can manipulate Eq.~\eqref{eq:2.4b} and \eqref{eq:2.4c}  in order to introduce the 
electron--phonon interaction in the basis of the phonon modes.  We analyze separately the first and second order terms.  

The first order can be written by using Eqs.~\eqref{eq:2.1}--\eqref{eq:2.4a} as:
\begin{multline}
\Delta\widehat{H}^{\(1\)}\(\RR\)=
\sum_{\qq\gl} 
\Bigg\{\[\sum_{\kk} \int_0 d\rr
\hat{\psi}^{\dagger}_{\kk}\(\rr\)
\xi_{\gql}\(\rr\)
\hat{\psi}_{\kk-\qq}\(\rr\)\] \\
+\Xi_{\qq\gl}\Bigg\}
 \(\hat{b}^{\dagger}_{-\qq \gl}+\hat{b}_{\qq \gl}\),
\label{eq:2.5}
\end{multline}
with
\begin{align}
\xi_{\gql}\(\rr\)=\partial_{\(\qq\gl\)} W_{e-n}\(\rr,\RR\).
\label{eq:2.5a}
\end{align}
The function $\partial_{\(\qq\gl\)} W_{e-n}\(\rr,\RR\)$ represents the derivative of the electron--phonon interaction along the phonon mode
$\(\qq,\gl\)$. The definitions of the $\partial_{\(\qq\gl\)}$ operator and the $\xi_{\gql}\(\rr\)$ function are given in the appendix~\ref{appB}. 
Note that, in Eq.~\eqref{eq:2.5}, the real--space integral is performed in the unit cell $\int_0$ and not in the whole crystal. This is because the sum running on all unit cell replicas
has been used to impose the momentum conservation at each vertex of the interaction terms (see Eq.~\eqref{eq:B.11}).

A similar derivation can be done for $\Xi_{\qq\gl}$, which represents the first-order derivative of the nucleus--nucleus potential 
\begin{align}
\Xi_{\gql}= \partial_{\(\qq\gl\)} W_{n-n}\(\RR\).
\label{eq:2.5b}
\end{align}
The second--order terms can be worked out in a similar way leading to the final form for their contribution to the electron--phonon interaction Hamiltonian:
\begin{multline}
\Delta\widehat{H}^{\(2\)}\(\RR\)=\\
\sum_{\qq\gl,\qq'\gl'} 
\Bigg\{\[\sum_{\kk} 
\int_0 d\rr
\hat{\psi}^{\dagger}_{\kk}\(\rr\)
\gt_{\gql,\qq'\gl'}\(\rr\)
\hat{\psi}_{\kk-\qq-\qq'}\(\rr\)\]\\
+\Theta_{\qq\gl,\qq'\gl'}\Bigg\} \(\hat{b}^{\dagger}_{-\qq \gl}+\hat{b}_{\qq \gl}\)
 \(\hat{b}^{\dagger}_{-\qq' \gl'}+\hat{b}_{\qq' \gl'}\).
\label{eq:2.6}
\end{multline}
In Eq.~\eqref{eq:2.6} we have introduced the functions $\gt$ and $\Theta$ whose definition is quite similar to Eq.~\eqref{eq:2.5a} and 
Eq.~\eqref{eq:2.5b}
\begin{align}
\theta_{\gql,\qq'\gl'}\(\rr\)= \frac{1}{2} \partial^2_{\(\qq\gl\)\(\qq'\gl'\)} W_{e-n}\(\rr,\RR\),
\label{eq:2.7a}
\end{align}
\begin{multline}
\Theta_{\gql,\qq'\gl'}=
 \frac{1}{2}\partial^2_{\(\qq\gl\)\(\qq'\gl'\)} W_{n-n}\(\RR\) \\
  -\Delta W_{n-n}^{ref}\(\RR\)\Big|_{\(\qq\gl\)\(\qq'\gl'\)},
\label{eq:2.7b}
\end{multline}
with $\left.\Delta W_{n-n}^{ref}\(\RR\)\right|_{\(\qq\gl\)\(\qq'\gl'\)}$ the reference nucleus--nucleus potential written in the 
phonon modes basis by plugging Eq.~\eqref{eq:2.1} into Eq.~\eqref{eq:ip.4}.

The explicit expressions for $\theta$ and for $\partial^2_{\(\qq\gl\)\(\qq'\gl'\)} W_{n-n}\(\RR\) $ are given in Appendix~\ref{appB}, Eqs.~\eqref{eq:B.15} and \eqref{eq:B.16}.

\begin{figure}[H]
\parbox[c]{4cm}{
\begin{center}
\epsfig{figure=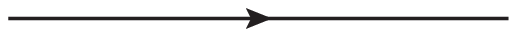,width=3cm}\\ %Diagrams/electron.eps
$G_{\kk}^{(0)}\(t\)$\\
(\ref{fig:1}.a)
\end{center}
}
\parbox[c]{4cm}{
\begin{center}
\epsfig{figure=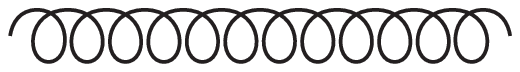,width=3cm}\\ %Diagrams/phonon.eps
$D_{\qq \gl}^{(0)}\(t\)$\\
(\ref{fig:1}.b)
\end{center}
}\\
\begin{center}
\parbox[c]{4cm}{
\begin{center}
\epsfig{figure=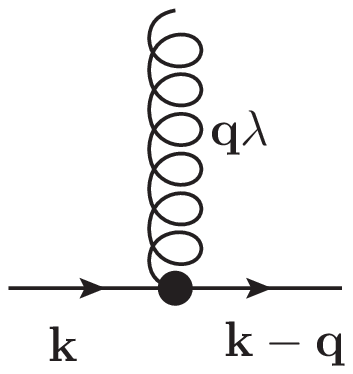,height=2cm}\\ %Diagrams/W_ei_1st_order.eps
$\bullet=\xi_{\qq\gl}\(\rr\)$\\
(\ref{fig:1}.c)
\end{center}
}
\parbox[c]{4cm}{
\begin{center}
\epsfig{figure=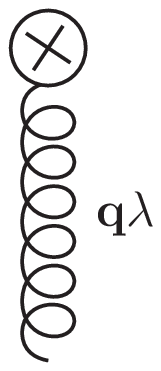,height=2cm}\\ %Diagrams/W_ii_1st_order.eps
$\bigotimes=\Xi_{\qq\gl}$\\
(\ref{fig:1}.d)
\end{center}
}\\
\parbox[c]{4cm}{
\begin{center}
\epsfig{figure=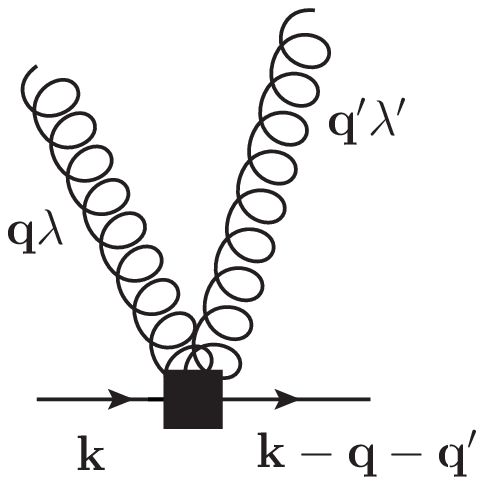,height=2cm}\\ %Diagrams/W_ei_2nd_order.eps
$\blacksquare=\theta_{\gql,\qq'\gl'}\(\rr\)$\\
(\ref{fig:1}.e)
\end{center}
}
\parbox[c]{4cm}{
\begin{center}
\epsfig{figure=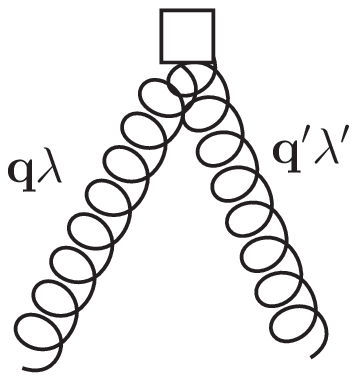,height=2cm}\\ %Diagrams/W_ii_2nd_order.eps
$\Box=\Theta_{\gql,\qq'\gl'}$\\
(\ref{fig:1}.f)
\end{center}
}
\end{center}
\caption{\footnotesize{
Diagrammatic representations of the electron and phonon propagators (diagrams (a) and (b)) and of
the first ($\xi$, diagram (c) and $\Xi$, diagram (d)) and second ($\theta$, diagram (e) and $\Theta$, diagram (f)) order interaction terms in the Taylor expansion of $\widehat{H}$ 
in powers of the nuclear displacements.  All interaction terms are written in the basis of the phonon displacements.
More definitions can be found in the text. 
}}
\label{fig:1}
\end{figure}

Finally, the total Hamiltonian $\widehat{H}\(\RR\)$, up to second order in the nuclear displacements, can be written as
\begin{widetext}
\begin{multline}
\widehat{H}\(\RR\)=\sum_{\nk} \gee_{\nk} \hat{c}^{\dagger}_{\nk} \hat{c}_{\nk}+\sum_{\qq\gl}\go_{\qq\gl}\(\hat{b}^{\dagger}_{\qq \gl}\hat{b}_{\qq \gl}+\frac{1}{2}\)+\frac{1}{2}\int\,d\rr_1 d\rr_2\, \hat{\psi}^{\dagger}\(\rr_1\) \hat{\psi}^{\dagger}\(\rr_2\) v\(\rr_1-\rr_2\) \hat{\psi}\(\rr_2\)\hat{\psi}\(\rr_1\)+\\
+\sum_{\qq\gl}\[ \int_0 d\rr \hat{\psi}^{\dagger}_{\kk}\(\rr\)
\xi_{\gql}\(\rr\)\hat{\psi}_{\kk-\qq}\(\rr\)+ \Xi_{\qq\gl}\]
\(\hat{b}^{\dagger}_{-\qq \gl}+\hat{b}_{\qq \gl}\) 
+\sum_{\qq\gl,\qq'\gl'}\Biggl[
\int_0 d\rr
\hat{\psi}^{\dagger}_{\kk}\(\rr\)
\gt_{\gql,\qq'\gl'}\(\rr\)
\hat{\psi}_{\kk-\qq-\qq'}\(\rr\)\\
+ \Theta_{\qq\gl,\qq'\gl'}\Biggr]  \(\hat{b}^{\dagger}_{-\qq \gl}+\hat{b}_{\qq \gl}\)
\(\hat{b}^{\dagger}_{-\qq' \gl'}+\hat{b}_{\qq' \gl'}\).
\label{eq:2.8}
\end{multline}
\end{widetext}
The diagrammatic transposition of the four electron--phonon interaction terms in Eq.~\eqref{eq:2.8} is given in Fig.~\eqref{fig:1}. We see that we have two terms with
electronic legs ($\bullet$ and $\blacksquare$). Those give direct contributions to the electronic propagator. 
In addition, there are two purely nuclear contributions ($\bigotimes$ and $\Box$) that do not contribute directly to the electron propagator but
can be combined with the two electronic interactions and still contribute to the electronic self--energy, as it will be clear in the following.
Those are commonly neglected in textbook theories of the electron--phonon interaction.
But from Eq.~\eqref{eq:2.5b} and Eq.~\eqref{eq:2.7b} we see that there is no reason, {\em a priori}, to assume that both
$\Xi_{\gql}$ and $\Theta_{\gql}$ are zero.

The $\Xi_{\gql}$ and $\Theta_{\gql}$ interaction terms do not have electronic legs because they arise from the purely nuclear potential ($W_{n-n}\(\RR\)$). Nevertheless, at it is 
evident from the above discussion, they can exchange momentum with the electronic subsytem. Energy, instead, is not exchanged as the nuclear potential is a static function. The momentum exchange 
reflects the change in the nuclear--nuclear potential induced by a nuclear displacement. This term is neglected in the jellium model because, as it will be clear 
in the following, the only allowed modes are acoustic excitations for which the zero frequency limit corresponds to the zero momentum limit. This contribution vanishes
as explained in Sec.\ref{sec:tad_pole}.

Thus, any coherent and accurate framework that aims at providing a consistent way of introducing screening and correlation in the perturbative expansion of
Eq.~\eqref{eq:2.8} will also have to answer to the key question about the role played by the nucleus--nucleus interaction.

%%%%%%%%%%%%%%%%%%%%%%%%%%%%%%%%%%%%
\section{The perturbative expansion}
\label{sec:the_expansion}
%%%%%%%%%%%%%%%%%%%%%%%%%%%%%%%%%%%%
Now that the total Hamiltonian has been split in the bare Hamiltonian, Eq.~\eqref{eq:2.3}, and in the interaction contributions, 
$\whW_{e-e}$, $\Delta\widehat{H}^{\(1\)}\(\RR\)$ (Eq.~\eqref{eq:2.5}),
and $\Delta\widehat{H}^{\(2\)}\(\RR\)$ (Eq.~\eqref{eq:2.6}), it is possible to perform a standard diagrammatic
analysis. In the following, we will work in the finite temperature regime where
the electronic Green's function is defined~\cite{ALEXANDERL.FETTER1971} as
\begin{align}
G\(1,2\)\equiv - Tr \Bigl\{ \hat{\gr}\(\gb\) T_t\[\hat{\psi}\(1\)\hat{\psi}^{\dagger}\(2\)\] \Bigr\}.
\label{eq:3.3}
\end{align}
In Eq.~\eqref{eq:3.3} $\gb=\(k_B T\)^{-1}$ and $T$ is the temperature. $T_t$ represents the time ordering product,
the $\hat{c}_{n\kk}\(t\)$ operator is written in the Heisenberg representation,
$\hat{\gr}\(\gb\)=e^{-\gb\(\widehat{H}-\mu\widehat{N}\)}/Tr  \Bigl\{ e^{-\gb\(\widehat{H}-\mu\widehat{N}\)} \Bigr\}$, with $\mu$ the chemical potential
and $\widehat{N}$ the total electronic number operator. We have introduced global variables to represent space and time components $1\equiv\(\rr_1,t_1\)$.
The average spanned by the trace operator runs over all possible interacting states weighted by the density operator $\hat{\gr}$.

By using Eq.~\eqref{eq:2.1a}, we can expand the Green's function in the electronic basis defined by the reference Hamiltonian:
\begin{multline}
G\(1,2\)=\\
\frac{1}{N}\sum_{nn'\kk}\phi_{n\kk}\(\rr_1\) \phi^{*}_{n'\kk}\(\rr_2\)G_{nn'\kk}\(t_1-t_2\),
\label{eq:3.1}
\end{multline}
with
\begin{equation}
G_{nn'\kk}\(t\)\equiv - Tr \Bigl\{ \hat{\gr}\(\gb\) T_t\[\hat{c}_{n\kk}\(t\)\hat{c}^{\dagger}_{n'\kk}(0)\] \Bigr\}.
\label{eq:3.1a}
\end{equation}
The electronic Green's function can also be expressed in a matrix representation as
\begin{equation}
\[\GG_{\kk}\(1,2\)\]_{nn'}=\phi_{n\kk}\(\rr_1\) \phi^{*}_{n'\kk}\(\rr_2\){ G}_{nn'\kk}\(t_1-t_2\).
\label{eq:3.7a}
\end{equation}
We will later use the same generic representation of Eqs.~\eqref{eq:3.1},\eqref{eq:3.1a}  and~\eqref{eq:3.7a} for the self--energy operator.

The non--interacting electronic Green's function $G^{\(0\)}_{nn'\kk}\(t\)$ is diagonal in the band index and it reduces to a simple exponential\cite{ALEXANDERL.FETTER1971}
\begin{multline}
G^{\(0\)}_{nn'\kk}\(t\)\equiv \gd_{n n'} e^{-\(\gee_{n\kk}-\mu\)t}\\
\times\[f\(\gee_{n\kk}\)\gt\(-t\)-\(1-f\(\gee_{n\kk}\)\)\gt\(t\)\],
\label{eq:3.4}
\end{multline}
with $f\(\gee\)\equiv\(e^{\gb\(\gee-\mu\)}+1\)^{-1}$ the Fermi-Dirac distribution function.
A similar expression holds for the non--interacting phonon propagator defined similarly to Eq.~\eqref{eq:3.3}, with bosonic phonon $\(\hat{b}_{\qq\gl}+\hat{b}^{\dagger}_{-\qq\gl}\)$ operators replacing the 
electronic ones $\hat{c}$:
\begin{multline}
D^{\(0\)}_{\qq\gl}\(t\)\equiv -\[1+n\(\go_{\qq\gl}\)\]\Bigl[e^{-\go_{\qq\gl}t}\gt\(t\)+e^{\go_{\qq\gl}t} \gt\(-t\)\Bigr]\\
-n\(\go_{\qq\gl}\)\Bigl[e^{\go_{\qq\gl}t}\gt\(t\)+e^{-\go_{\qq\gl}t} \gt\(-t\)\Bigr],
\label{eq:3.4q}
\end{multline}
with $n\(\gee\)\equiv\(e^{\gb\gee}-1\)^{-1}$ the Bose-Einstein distribution function.
Thanks to the standard many-body approach and perturbative expansion,
it is possible to rewrite $\GG_{\kk}\(t\)$ in terms of $\GG_{\kk}^{\(0\)}\(t\)$ and the electronic self--energy operator $\Sigma_{\kk}$ by means of 
the Dyson equation~\cite{ALEXANDERL.FETTER1971}
\begin{multline}
 \GG_{\kk}\(1,2\)=\GG^{\(0\)}_{\kk}\(1,2\)+
\int\,d3 d4 \GG^{\(0\)}_{\kk}\(1,3\)\\
\times {\bf \Sigma}_{\kk}\(3,4\) \GG_{\kk}\(4,2\),
\label{eq:3.7}
\end{multline}
where the $\Sigma_{\kk}$ matrix has been introduced following a definition similar to Eq.~\eqref{eq:3.7a}. Eq.~\eqref{eq:3.7} is written in 
diagrammatic form in Fig.(\ref{fig:2}.a).

Note that in Eq.~\eqref{eq:3.7} the time variables $t_3$ and $t_4$ runs in the range $ \[0,\gb\] $.

In the following subsections we will write $\Sigma$ using approximations with an increasing level of correlation, self--consistency and screening in order to investigate the effect of the different \ep\,interaction terms. The solution of the Dyson equation
corresponds to an infinite series in terms of the Green's function and the self-energy.
Consistently with the harmonic approximation (the expansion with respect to the nuclear displacements is limited to the
second power in Eqs.~\eqref{eq:1.4} and~\eqref{eq:2.8}), we will work
up to second order with respect to $\Delta\widehat{H}^{\(1\)}\(\RR\)$, and only to first order in $\Delta\widehat{H}^{\(2\)}\(\RR\)$. 
Higher orders of nuclear displacements
might appear as a consequence of self-consistency or screening (in the Dyson equation), but we will consider them to be negligible or to have no impact, 
consistently with our choice of the harmonic approximation.
By contrast, for the electron-electron interaction, there will be no such approximation: higher-order powers of the electron-electron interaction will be significant. 

%%%%%%%%%%%%%%%%%%%%%%%%%%%%%%%%%%%%%%%%%%%%%%%%%%%%%%%%%%%%%%%%%%%%%%%%%%%%%%%%%%
\subsection{The Hartree, Debye--Waller and tad--poles self--energies}
\label{sec:lowest_order}
\begin{figure}[H]
\parbox[c]{8cm}{
\begin{center}
\epsfig{figure=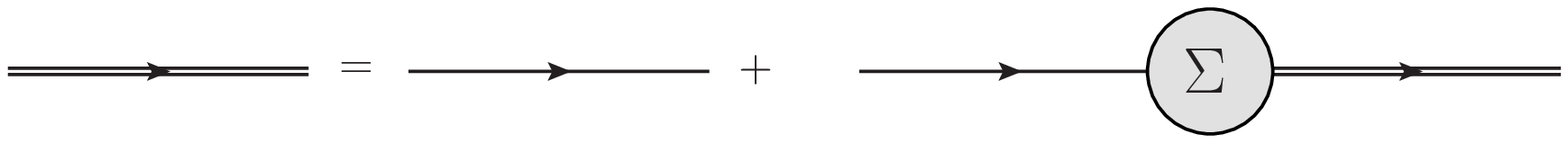,width=6cm}\\ %Diagrams/dyson_non_local.eps
(\ref{fig:2}.a)
\end{center}
}\\
\parbox[c]{8cm}{
\begin{center}
\epsfig{figure=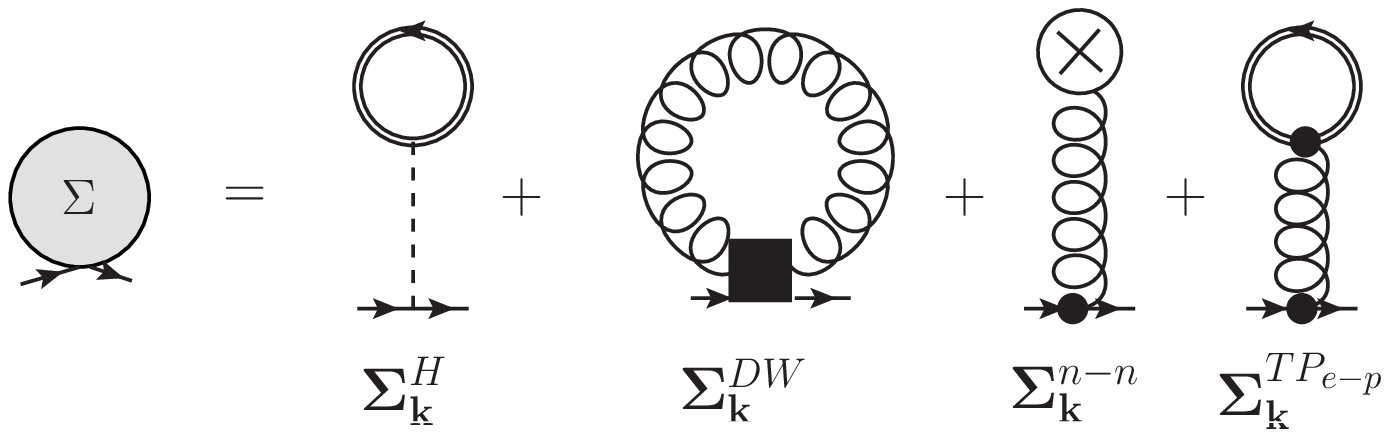,width=6cm}\\ %Diagrams/sigma_lowest_order.eps
(\ref{fig:2}.b)
\end{center}
}
\caption{\footnotesize{
Dyson equation written in terms of diagrams (frame\,(a)). In frame\,(b), instead,
the lowest-order electronic self--energy in the electron--electron and electron--phonon interaction is shown. We see the usual electronic Hartree contribution (first diagram) $\bgS_{\kk}^{H}$,   the well-known
Debye--Waller term (second diagram) $\bgS_{\kk}^{DW}$, and the electron--phonon induced tad--pole (fourth diagram) $\bgS_{\kk}^{TP_{e-n}}$. In addition we see the
appearance of a new diagram due to the derivative of the nucleus--nucleus potential (third diagram),  $\bgS_{\kk}^{n-n}$. This diagram plays a crucial role
in balancing the contribution from $\bgS_{\kk}^{TP_{e-n}}$ that, indeed, is not zero in general. 
}}
\label{fig:2}
\end{figure}

%%%%%%%%%%%%%%%%%%%%%%%%%%%%%%%%%%%%%%%%%%%%%%%%%%%%%%%%%%%%%%%%%%%%%%%%%%%%%%%%%%
We analyze first the electronic self-energy that is obtained by considering only one interaction node attached to the electronic Green's function, and select the lowest 
non-vanishing  diagrams. This self--energy can be obtained mathematically from the Feynmann diagrams in figure~\ref{fig:2}.b, following the diagrammatic rules of Ref.~\onlinecite{ALEXANDERL.FETTER1971}, for example. 
It is composed of four terms: the Hartree $\Sigma^{H}$, Debye-Waller $\Sigma^{DW}$, nucleus-nucleus $\Sigma^{n-n}$ and electron-nucleus $\Sigma^{e-n}$ self-energy:
\begin{gather}
\Sigma^{H}\(1\)= \int\,d2 v\(1,2\)G\(2,2^+\),
\label{eq:3.7d}\\
\Sigma^{DW}\(1\)= -\sum_{\qq,\gl}\gt_{\qq\gl,-\qq\gl}\(1\) D^{\(0\)}_{\qq\gl}\(0^{-}\),
\label{eq:3.7e}\\
\Sigma^{n-n}\(1\)= \limq\[\sum_{\gl} \xi_{\qq\gl}\(1\)D^{\(0\)}_{\qq\gl}\(0^-\)\Xi^{*}_{\qq\gl}\],
\label{eq:3.7f}
\end{gather}
where $v\(1,2\)\equiv \gt\(t_1-t_2\)v\(\rr_1-\rr_2\)$.
The \ep\, induced tad--pole contribution is
\begin{multline}
\Sigma^{TP_{e-n}}\(1\)=\limq\Bigg\{\sum_{\gl} \xi_{\qq\gl}\(1\)D^{\(0\)}_{\qq\gl}\(0^-\)\\
\times\[\int\,d2 \xi^{*}_{\qq\gl}\(2\)G\(2,2^+\)\]\Bigg\}.
\label{eq:3.7g}
\end{multline}
The other diagram (Fan diagram) appearing at second order in $\Delta\widehat{H}^{\(1\)}\(\RR\)$ will be analyzed in the next subsection, Sec.~\ref{sec:HF}, 
together with the Fock diagram from $\whW_{e-e}$. They both have two interactions nodes attached to the electronic Green's function.

The self-energy $\Sigma^{H}$ is the usual Hartree contribution which is a tad--pole diagram made of a Coulomb interaction that connects the incoming electronic propagator with another, closed,
electronic loop. Also the $\Sigma^{TP_{e-n}}$ is a tad--pole diagram, but, in contrast to the Hartree term,
the interaction is not electronic but phonon mediated. 

The nucleus-nucleus self-energy $\Sigma^{n-n}$ is a new term that has never been discussed before in the literature. It comes from the merging of a nucleus--nucleus and electron--nucleus interaction. It acquires an electronic character thanks to the contraction with the electronic propagator embodied in the electron--nucleus interaction\,($\bullet$).

Finally $\Sigma^{DW}$ is the well--known Debye--Waller\,(DW) self--energy. It represents the lowest (first) order electronic self--energy in the second--order derivative of the Hamiltonian.

The total self--energy is local in time and space and therefore the Dyson equation of Eq.~\eqref{eq:3.7} reduces to 
\begin{multline}
 \GG_{\kk}\(1,2\)=\GG^{\(0\)}_{\kk}\(1,2\)\\
 +\int\,d3 \GG^{\(0\)}_{\kk}\(1,3\) {\bf \Sigma}_{\kk}\(3\) \GG_{\kk}\(3,2\),
\label{eq:3.7b}
\end{multline}
with
\begin{multline}
 {\bf \Sigma}_{\kk}\(1\)=\\
 {\bf \Sigma}^{H}_{\kk}\(1\)+ {\bf \Sigma}^{DW}_{\kk}\(1\)+  {\bf \Sigma}^{n-n}_{\kk}\(1\)+ {\bf \Sigma}^{TP_{e-n}}_{\kk}\(1\).
\label{eq:3.7c}
\end{multline}

The different contributions shown in Eq.~\eqref{eq:3.7c} have the following properties:
\begin{itemize}
\item[(a)] All self--energy contributions are bare. No screening is present. This is not what should be applied for practical calculations because, as it will be clear
in the following, bare potentials lead also to unphysical properties. 
Moreover from DFPT, we know that ${\bf \Sigma}^{DW}$ is screened. 
In the original work of AHC this screening was introduced in a semi--empirical manner while in the more advanced 
approach based on DFPT\cite{baroni2001,Gonze1995} the electron--nucleus interaction is screened in the self--consistent solution of the Kohn--Sham equation.
It is clear, however, that from a rigorous MB approach this screening is not present in the original Hamiltonian and must be build--up by the electronic correlations. 
How does this screening emerge from a MB perspective?
\item[(b)] The nucleus-nucleus self-energy $\bgS_{\kk}^{n-n}$ is a new contribution that is not present in any treatment of the electron--phonon interaction where the nuclear density is approximated with an homogeneous charge density. In this work, the nuclear coordinates are instead coherently taken into account. This is an essential step to bridge the MBPT and DFT approaches.
\item[(c)] In the standard approach to the electron--phonon interaction the electron-nucleus self-energy $\bgS^{TP_{e-n}}$ is commonly neglected. However the arguments that motivate this approximation~\cite{ALEXANDERL.FETTER1971} are based
on two specific approximations: (i) the nuclear interaction is dressed and (ii) there are only acoustic modes. In general, however, any system has both acoustic
and optical modes and the $\bgS^{TP_{e-n}}$ self--energy is not vanishing. What is its role and is it justified to neglect it?
\end{itemize}
In order to answer those questions we proceed with a detailed analysis of the two series of diagrams connected with the dressing of the tad--pole and of the Debye--Waller terms.

%%%%%%%%%%%%%%%%%%%%%%%%%%%%%%%%%%%%%%%%%%%%%%%%%%%%%%%%%%%%%%%%%%%%%%%%%%%%%%%%%%%%%%%%%%%%%%%
\subsubsection{The electron--phonon induced tad--pole diagram and the nucleus--nucleus interaction contribution}
\label{sec:tad_pole}
%%%%%%%%%%%%%%%%%%%%%%%%%%%%%%%%%%%%%%%%%%%%%%%%%%%%%%%%%%%%%%%%%%%%%%%%%%%%%%%%%%%%%%%%%%%%%%%
The sum of the $\Sigma^{n-n}$ and $\Sigma^{TP_{e-n}}$ is
\begin{multline}
\Sigma^{n-n}\(1\)+\Sigma^{TP_{e-n}}\(1\)=
\limq\Bigg\{\sum_{\gl} \xi_{\qq\gl}\(1\)D^{\(0\)}_{\qq\gl}\(0^-\)\\
\times \Bigl[\Xi^{*}_{\qq\gl}+
\int\,d2 \xi^{*}_{\qq\gl}\(2\)G\(2,2^+\)\Bigr]\Bigg\}.
\label{eq:3.8}
\end{multline}
This sum would be zero if the $\xi_{\qq\gl}\(1\)$ prefactor or if
the expression between brackets vanishes. 
However the derivative of the bare ionic potential, when $\qq\rightarrow{\bf 0}$, diverges
like $|\qq|^{-1}$. Thus Eq.\eqref{eq:3.8} is, actually, divergent.
In the Jellium model  the
screening~\cite{ALEXANDERL.FETTER1971} of the electron--nucleus potential regularizes this divergence and
the {\em dressed} e--n interaction vanishes when $\qq\rightarrow{\bf 0}$. This is the standard motivation
used to neglect the contribution coming from the integral appering in the r.h.s of  Eq.\eqref{eq:3.8}. 
The term due to $\Xi$, instead, has been never considered before.

We focus our attention, instead, on the sum of the terms between brackets appearing in Eq.~\eqref{eq:3.8} for small but not vanishing values of $\qq$. If we work it out we can rewrite it in a more clear way.
We notice that $ G\(2,2^+\)=\gr\(2\)$ and, from Eq.~\eqref{eq:B.12} and Eq.~\eqref{eq:B.15}, we have that
\begin{widetext}
\begin{multline}
 \Xi^{*}_{\qq\gl}+\int\,d\rr_2\xi^{*}_{\qq\gl}\(\rr_2,\RR\)G\(2,2^+\)=
\sum_{ls\ga}
\frac{\eta_{\ga}\(\qq \gl|s\) e^{-i\qq \cdot \overline{\RR}_{ls}}}{\sqrt{2NM_s\go_{\qq\gl}}} 
\overline{\partial_{R_{ls\ga}} 
\[\int\,d\rr_2 W_{e-n}\(\rr_2,\RR\)\gr\(\rr_2\)+ W_{n-n}\(\RR\)\]}.
\label{eq:3.8a}
\end{multline}
\end{widetext}
Before proceeding in the evaluation of Eq.\eqref{eq:3.8a} we notice that from the Dyson equation (Eq.~\eqref{eq:3.7}) it follows that
% within the harmonic approximation, we can approximate
%$\RR=\overline{\RR}$ in Eq.\eqref{eq:3.8a}. $\overline{\RR}$ are the atomic positions calculated without taking into account the effect of the \ep\,interaction. This 
%fact can be proved by noticing that, from the Dyson equation Eq.~\eqref{eq:3.7}, it follows that
\begin{align}
\GG_{\kk}\approx \GG^{\(0\)}_{\kk}+\Delta\GG^{\(e-e\)}_{\kk}+O\(\dRu^2\),
\label{eq:3.10c}
\end{align}
with $\Delta\GG^{\(e-e\)}$ the change in the Green's function due to el--el correlation effects. Eq.~\eqref{eq:3.10c} implies that
$\gr\approx \gr^{\(e-e\)}+O\(\dRu^2\)$ with $ \gr^{\(e-e\)}$ the exact charge of the system with atoms frozen at the equilibrium positions. 
Therefore, as in Eq.~\eqref{eq:3.8a} $\partial_{\RR_s} W_{e-n}\(\rr,\RR\)\propto \dRu$, it follows that, within the harmonic approximation, we can safely assume
$\gr\sim\gr^{\(e-e\)}$.
% in Eq.~\eqref{eq:3.8a}.

It follows than, that within the harmonic approximation, 
the overlined quantity in  Eq.~\eqref{eq:3.8a} is minus the force $\FF_s$ acting on the nucleus located at $\overline{\RR}$ 
\begin{multline}
\FF_{ls}\equiv
- \partial_{\RR_{ls}}\Big[ \int\,d\rr_2 W_{e-n}\(\rr_2,\RR\)\gr^{\(e-e\)}\(\rr_2\)\\
+W_{n-n}\(\RR\)\Big]_{\RR=\oRR}.
\label{eq:3.10}
\end{multline}
By using the Hellmann--Feynman theorem it follows that
\begin{align}
\FF_{ls}\equiv -\partial_{\RR_{ls}}\la \Psi_0 | \widehat{H}_{BO}\(\RR\)|  \Psi_0\ra_{\RR=\oRR},
\label{eq:3.10a}
\end{align}
with $|\Psi_0\ra$ the {\em exact} electronic ground state of the total frozen Hamiltonian, $\widehat{H}\(\overline{\RR}\)$. 

%The Taylor expansion of Eq.~\eqref{eq:1.4}  is done around the equilibrium positions $\(\overline{\RR}_{ls}\)$ defined by the condition that
%$\FF_s=\sum_l \FF_{ls}=0$ for all nuclei. However in Eq.~\eqref{eq:3.10} $\gr$ is the finite temperature ground--state density of the Born-Oppenheimer Hamiltonian. 
%This follows from the fact
%that the ground--state $|\Psi_0\ra$ is the finite temperature electronic ground state and, as a consequence, the  positions $\(\RR\)$ for which the condition $\FF_s={\bf 0}$ must be
%interpreted as relative to the finite temperature electronic system. 
%Physically this means that, for example, when the temperature is increased the equilibrium positions will change and the system volume as well, 
%due to the modification of bonding. Still, this is a purely
%electronic effect, not due to the anharmonic phonon-phonon interactions or the quasi-harmonic thermal expansion. 

Thus we can draw the following conclusion: {\em if the self--energy is chosen in such a way that the nuclear positions and density correspond to the exact
electronic ground state 
then it follows}
\begin{align}
\Sigma^{TP_{e-n}}\(1\)+\Sigma^{n-n}\(1\)=0.
\label{eq:3.10b}
\end{align}
This is true for the exact self--energy but it is not true for any approximation of the self--energy unless the Born-Oppenheimer energy of the
system is calculated accordingly by means of MBPT (for example by using the Luttinger--Ward expressions~\cite{ALEXANDERL.FETTER1971}).

The condition represented by Eq.~\eqref{eq:3.10b} reveals the crucial role played by the nucleus--nucleus interaction. It is only thanks to the
coherent inclusion of electron-nucleus and nucleus-nucleus contributions that the theoretical framework can lead to the justification
of the AHC approach or to the textbooks results. A formal condition for the
tad--pole diagram to be zero can therefore be defined.  

As an additional approximation we notice that one of the most widely approximation used in the litterature  is to  treat correlation effects non self--consistently. In pratice this means to use the
Dyson equation to renormalize the single particle energies but not the wave--functions. As a consequence, within this approximation, the charge density is assumed to be well described by the 
one calculated within DFT. This approximation has an important and usefull consequence. Eq.~\eqref{eq:3.10c} would impose to use as ionic positions ($\overline\RR$)
the ones calculated with a level of correlation coherent with the one introduced in $\gr^{\(e-e\)}$. As this is a hardly (if not impossible) to
do in practice the use of the DFT charge allows to approximate both $\gr\sim\gr_{\(0\)}$ and $\overline\RR\approx\overline\RR^{\(0\)}$  in Eq.~\eqref{eq:3.10}.
In this case $\RR^{\(0\)}$ are the DFT equilibrium atomic positions that are a simple by--product of any DFT calculation.

%%%%%%%%%%%%%%%%%%%%%%%%%%%%%%%%%%%%%%%%%%%%%%%%%%%%%%%%%%%%%%%%%%%%%%%%%%%%%%%%%%%%%%%%%%%%%%%%%%%%
\subsubsection{Self--consistent diagrams: dressing of the internal Green's functions and of the bare interactions}
\label{sec:linearization}
%%%%%%%%%%%%%%%%%%%%%%%%%%%%%%%%%%%%%%%%%%%%%%%%%%%%%%%%%%%%%%%%%%%%%%%%%%%%%%%%%%%%%%%%%%%%%%%%%%%%
Before starting the analysis of the diagrammatic structure of the self--energy we notice that at any order
of the diagrammatic expansion we can clearly distinguish between diagrams that dress the internal 
electronic propagators and the interaction. 

\begin{figure}[H]
\begin{center}
\epsfig{figure=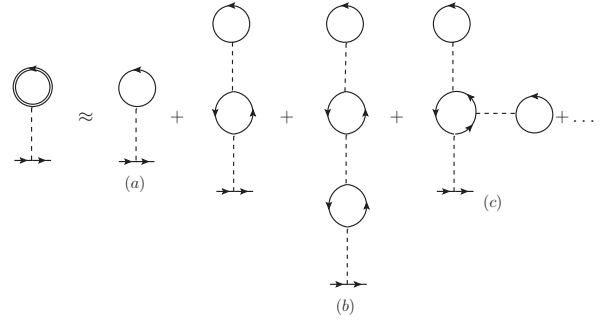,width=8cm}\\ %Diagrams/sigma_hartree_expansion.eps
\end{center}
\caption{\footnotesize{Diagrammatic expansion of the self--consistent Hartree self--energy. Diagram (a) is the non self--consistent contribution corresponding to the
bare electronic charge. Diagram (b) is composed of bare bubble diagrams.  These diagrams belongs to the family of diagrams that dress the interaction. Diagram (c), instead, represents a dressing of the internal electronic propagator. All diagrams of this kind
can be, within the quasi--particle approximations, summed in a definition of a new independent particle Hamiltonian, as explained in the text.
}}
\label{fig:3}
\end{figure}

A clear definition of these two families of diagrams can be done 
by using the simple Hartree approximation for the self--energy. Its diagrammatic expansion is shown in Fig.\eqref{fig:3} and we notice that at third order, two different diagrams appear ($(b)$ and $(c)$). In the case of diagram $(b)$ the interaction builds up a series of bubbles that describes electron--hole pair excitations. These bubbles, when summed to all orders,
reduce to the well--known Random Phase Approximation for the response function as it will be shown explicitly in Sec.~\eqref{sec:DW}. 

The diagram $(c)$, instead, represents a bare self--energy insertion in an internal Green's function. Any other diagram where an internal propagator is dressed belongs to this family. The effect of these diagrams is to renormalize the single particle states. This can be easily visualized in the quasi--particle approximation where all internal propagator self--interaction contributions can be summed
in the definition of a new set of independent particle energies, $\{\gee_{\nk} \}$.

In this work, we are interested in building up a scheme to concretely compare the Many--Body and DFT schemes as far as the e--n interaction is concerned. In order to greatly simplify the analysis, we will focus on the first series of diagrams disregarding all diagrams that correspond to a dressing of internal electronic propagators. In the self--consistent Hartree case this amounts to neglect
the diagram $(c)$ and, in practice, this means that the screening of the interaction is described by oscillations (described by bubble diagrams)  of the bare charge. This approximation is commonly
used in the \aimbpt scheme and can be written analytically as
\begin{align}
\GG^{internal}_{\kk}\approx \GG^{\(0\)}_{\kk}.
\label{eq:3.10d}
\end{align}
When applied in the diagramatic context we will refer to Eq.\eqref{eq:3.10d} as the {\em linearization procedure}. The reason of the name will be clear in the next section.

%%%%%%%%%%%%%%%%%%%%%%%%%%%%%%%%%%%%%%%%%%%%%%%%%%%%%%%%%%%%%%%%%%%%%%%%%%%%%%%%%%%%%%%%%%%%%%%
\subsubsection{Screening of the second--order \ep interaction and of the Debye--Waller diagram}
\label{sec:DW}
%%%%%%%%%%%%%%%%%%%%%%%%%%%%%%%%%%%%%%%%%%%%%%%%%%%%%%%%%%%%%%%%%%%%%%%%%%%%%%%%%%%%%%%%%%%%%%%
As mentioned earlier, one important aspect that must be included in the perturbative analysis in order to bridge it with 
the DFPT formalism is the screening of the electron--phonon interaction terms. 
How does screening build up\,? This question could appear easy to answer as the series of diagrams that screen the lowest order ($\xi$) interaction is, indeed,
well known in the litterature. But, what about the second
order interaction, $\Xi$\,?

In the original AHC work, the DW self--energy is written, from the beginning, in terms of a statically screened $W_{e-n}$ interaction. 
However this screening cannot be introduced directly in the Hamiltonian. It must appear
as a result of the diagrammatic expansion.

In addition, from the discussion of the previous section, it follows that $\Sigma^{TP_{e-n}}\(1\)+\Sigma^{n-n}\(1\)$ is not zero for any self--energy that does not
reproduce the exact reference density and the exact corresponding nuclear positions. As a matter of fact this is a condition hard to fulfill in any practical implementation as
it is computationally very difficult to find the nuclear positions corresponding to a specific level of approximation for $\Sigma$. 

Even if the condition given by Eq.~\eqref{eq:3.10b} can be used as a simple approximation it is instructive, for the moment, to keep the two self--energies
in our derivation in order to see their effect on the definition of the screening. 

\begin{figure}[H]
\begin{center}
\epsfig{figure=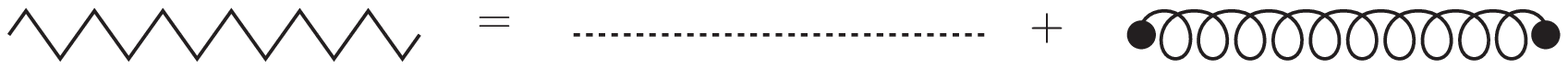,width=8cm} %Diagrams/zigzag
\end{center}
\caption{\footnotesize{
The total bare interaction $W_{\qq}^{\gl}\(1,2\)$.
}}
\label{fig:4}
\end{figure}

At the lowest order of the perturbative expansion the self--energy shown in Diagram~\ref{fig:2}.b can be explicitly written as
\begin{multline}
\bgS_{\kk}\(1\)= \sum_{\gl}\limq\Bigg\{\int\,d2 W_{\qq\gl}\(1,2\) \GG_{\kk}\(2,2^+\)\\
+\xi_{\qq\gl}\(1\)D^{\(0\)}_{\qq\gl}\(0^{-}\)\Xi^{*}_{\qq\gl}\\
-\sum_{\qq}\gt_{\qq\gl,-\qq\gl}\(1\) D^{\(0\)}_{\qq\gl}\(0^{-}\)\Bigg\},
\label{eq:3.11}
\end{multline}
where we have introduced a total bare electron--electron interaction $W_{\qq\gl}\(1,2\)$ (see Fig.~\eqref{fig:4}) defined as
\begin{multline}
W_{\qq\gl}\(1,2\)\equiv v_{\qq}\(1,2\)\gd\(t_1-t_2\)\\
+\sum_{\gl}\xi_{\qq\gl}\(1\)
D^{\(0\)}_{\qq\gl}\(t_1-t_2\)\xi^{*}_{\qq\gl}\(2\),
\label{eq:3.11a}
\end{multline}
with $v_{\qq}\(1,2\)$ the periodic $\qq$--component of the Fourier transformation of the bare Coulomb interaction
\begin{align}
v\(1,2\)=\int_{BZ}\,\frac{d\qq}{\(2\pi\)^3} e^{i\qq\cdot\(\rr_1-\rr_2\)} v_{\qq}\(1,2\),
\end{align}
with the integral restricted to the Brillouin Zone\,(BZ) only. 

\begin{figure}[H]
\parbox[c]{8cm}{
\begin{center}
\epsfig{figure=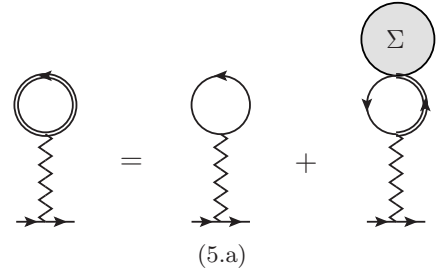,width=6cm}\\ %Diagrams/sigma_h_2.eps
(\ref{fig:5}.a)
\end{center}
}
\parbox[c]{8cm}{
\begin{center}
\epsfig{figure=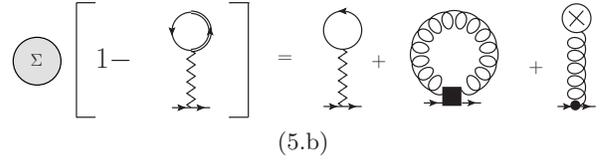,width=8cm}\\ %Diagrams/sigma_h_3.eps
(\ref{fig:5}.b)
\end{center}
}
\parbox[c]{8cm}{
\begin{center}
\epsfig{figure=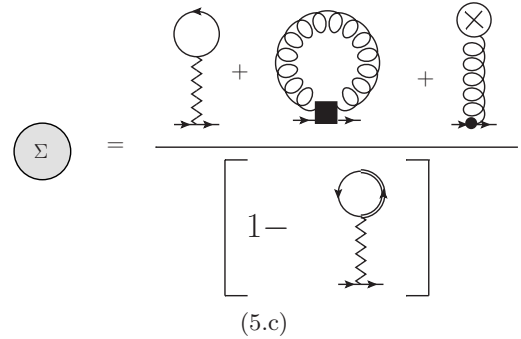,width=7cm}\\ %Diagrams/sigma_h_4.eps
(\ref{fig:5}.c)
\end{center}
}
\caption{\footnotesize{
Diagrammatic proof that self--consistency at the Hartree level is equivalent to screening, at the 
time--dependent Hartree level. The proof is obtained by
using the Dyson equation inside the definition of the Hartree self--energy. This can be solved in terms of the self--energy
itself by a simple Fourier transformation because the Hartree self--energy is local in time. The mathematical transposition of
this proof is discussed in the text.
}}
\label{fig:5}
\end{figure}

The total bare electron-electron interaction $W_{\qq\gl}$ gets dressed (i.e. screened) by self-consistency when the Green's function (Eq.~\eqref{eq:3.7b}) is inserted into the self--energy (Eq.~\eqref{eq:3.11}). We can also sum over $\kk$ as the Hartree self--energy 
is a local function and get
\begin{widetext}
\begin{multline}
\Sigma\(1\)=\frac{1}{N}\sum_{\kk}\bgS_{\kk}\(1\)=
\sum_{\gl}\bigg\{ \int\,d2 \limq\[W_{\qq \gl}\(1,2\)\] 
\times \[ G^{\(0\)}\(2,2^+\)+\int\,d3 G^{\(0\)}\(2,3\){\bf \Sigma}\(3\) G\(3,2\)\]\\
+\limq\[\xi_{\qq\gl}\(1\)D^{\(0\)}_{\qq\gl}\(0^{-}\)\Xi^{*}_{\qq\gl}\]
-\sum_{\qq}\gt_{\qq\gl,-\qq\gl}\(1\) D^{\(0\)}_{\qq\gl}\(0^{-}\)\bigg\}.
\label{eq:3.13}
\end{multline}
\end{widetext}
The $W_{\qq \gl}\(1,2\)\GG_{\kk}\(2,2^+\)$ term appearing in Eq.\eqref{eq:3.11} is written in terms of Feynman diagrams in Fig.(\ref{fig:5}.a).

Now we group all ${\bf \Sigma}$ operators to the left hand side of the equation.
We have that
\begin{widetext}
\begin{multline}
\int\,d3 \Sigma\(3\)\Bigl[ \gd\(1,3\) 
 -\sum_{\gl}\limq\[W_{\qq\gl}\(1,2\)\]  G^{\(0\)}\(2,3\)  G\(3,2\)\Bigr]=\\
\sum_{\gl}\bigg\{\int\,d2 \limq\[W_{\qq\gl}\(1,2\)\]  G^{\(0\)}\(2,2^+\)
+\limq\[\xi_{\qq\gl}\(1\)D^{\(0\)}_{\qq\gl}\(0^{-}\)\Xi^{*}_{\qq\gl}\]
-\sum_{\qq}\gt_{\qq\gl,-\qq\gl}\(1\) D^{\(0\)}_{\qq\gl}\(0^{-}\)\bigg\},
\label{eq:3.14}
\end{multline}
\end{widetext}
and the corresponding diagrams are shown on Fig.(\ref{fig:5}.b).

Now, Eq.~\eqref{eq:3.14} is not linear in the sense that the right-hand side depends on the dressed $G$ because of the perturbative expansion of
the inverse of the square bracket quantity appearing in the left-hand side.
By using the discussion of Sec.\ref{sec:linearization} we observe that all dressed
$G$'s are internal Green's functions. This can be deduced also by the expansion of the diagrammatic fraction appearing in Fig.(\ref{fig:5}.c). Then we can apply the {\em linearization procedure}, Eq.\eqref{eq:3.10d} to approximate $G\(3,2\)$ with $G^{\(0\)}\(3,2\)$ in the Eq.\eqref{eq:3.14}. This allows to define
the single--particle response function $\chi^{\(0\)}$
\begin{equation}
\limq {\bf\chi}_{\qq}^{\(0\)}\(1,2\)\equiv \frac{1}{N}\sum_{\kk} \GG_{\kk}^{\(0\)}\(1,2\) \GG^{\(0\)}_{\kk}\(2,1\).
\label{eq:3.9}
\end{equation}

Then, we define the dielectric matrix in the Time-Dependent Hartree approximation (see later) as
\begin{multline}
\gee_{\qq}^{tdh}\(1,2\)\equiv  \\
  \gd\(1,2\) - \sum_{\gl}\int\,d3  W_{\qq\gl}\(1,3\) \chi_{\qq}^{\(0\)}\(3,2\).
\label{eq:3.9p}
\end{multline}

By using Eq.~\eqref{eq:3.14} into Eq.~\eqref{eq:3.9} the screening of the bare potential $W$ and
electron-phonon terms appears
so that
\begin{multline}
\Sigma\(1\)=
\sum_{\gl}\bigg\{\int\,d2 \limq\[\widetilde{W}_{\qq\gl}\(1,2\)\]  G^{\(0\)}\(2,2^+\)\\
+\limq\[\widetilde{\xi}_{\qq\gl}\(1\)D^{\(0\)}_{\qq\gl}\(0^{-}\)\Xi^{*}_{\qq\gl}\]\\
-\sum_{\qq}\widetilde{\gt}_{\qq\gl,-\qq\gl}\(1\) D^{\(0\)}_{\qq\gl}\(0^{-}\)\bigg\}
\label{eq:3.15}
\end{multline}
with $\widetilde{W}$, $\widetilde{\xi}$ and $\widetilde{\gt}$ the dressed counterparts of the bare $W$, $\xi$ and $\gt$ functions
\begin{equation}
\widetilde{\xi}_{\qq\gl}\(1\)=\int d2  \[\gee_{\qq}^{tdh}\(1,2\)\]^{-1} \xi_{\qq\gl}\(\rr_2\)\label{eq:3.12},
\end{equation}
\begin{equation}\label{eq:3.12.1}
\widetilde{\gt}_{\qq\gl,-\qq\gl}\(1\)=
\int d2  \[\limq\gee_{\qq}^{tdh}\(1,2\)\]^{-1} \theta_{\qq\gl,-\qq\gl}\(\rr_2\),
\end{equation}
\begin{equation}\label{eq:3.12.2}
\widetilde{W}_{\qq}\(1,2\)=\int d3 \[\gee_{\qq}^{tdh}\(1,3\)\]^{-1} W_{\qq}\(3,2\).
\end{equation}
Eq.~\eqref{eq:3.15} and Eqs.~\eqref{eq:3.12}, \eqref{eq:3.12.1}, \eqref{eq:3.12.2} represent an important result of the present work. Indeed, they show that self--consistency screens
the interaction lines of all diagrams, including the Debye--Waller one. This result will be crucial in discussing how to include higher-order diagrams 
avoiding double-counting problems.

Indeed, there are two well known ways of increasing the order of the perturbative expansion. One is to add skeleton diagrams and the other is to use
self--consistency. This second path is extremely important as it provides the way for a given self--energy to fulfill conserving conditions\cite{Strinati1988}. 
Skeleton (as well as reducible) diagrams are known to build--up the screening of the electron--electron and electron--nucleus interactions. 
In this section we have shown that screening arises also from self--consistency. 

\begin{figure}[H]
\begin{center}
\parbox[c]{8cm}{
\begin{center}
\epsfig{figure=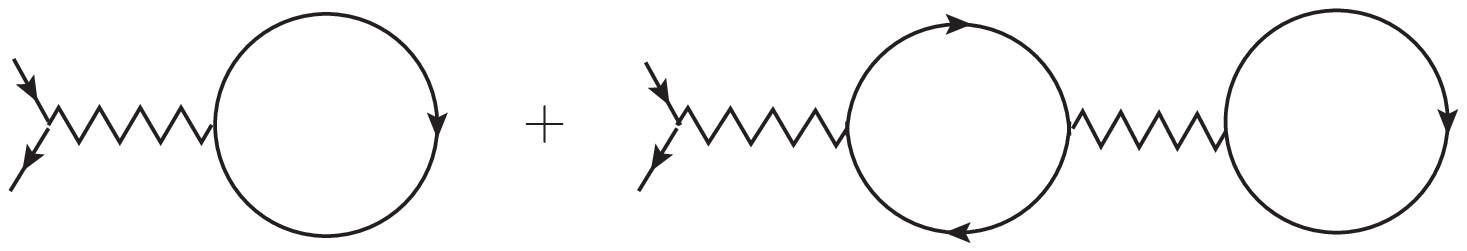,width=6cm}\\ %Diagrams/Hartree_tdh.eps
(\ref{fig:6}.a)
\end{center}
}
\parbox[c]{8cm}{
\begin{center}
\epsfig{figure=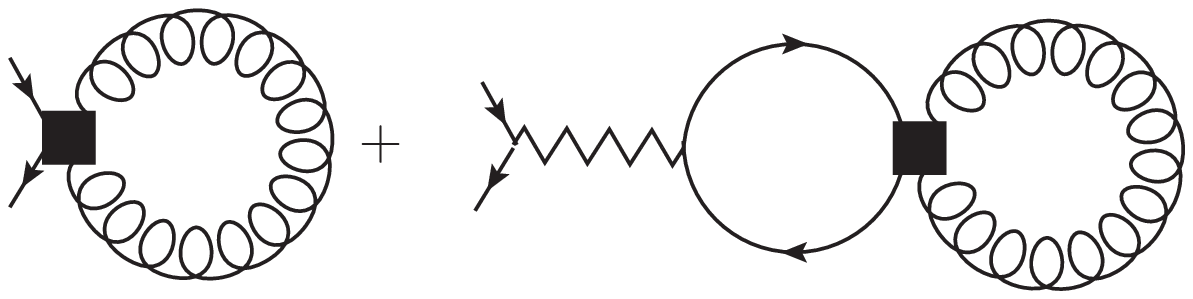,width=6cm}\\ %Diagrams/DW_tdh.eps
(\ref{fig:6}.b)
\end{center}
}
\end{center}
\caption{\footnotesize{
First two diagrams contributing to the screening of the Hartree (upper diagram) and DW (lower diagram) term. The screening can be written as the action of a 
time--dependent Hartree screening function (see text). The dielectric function contains only the Hartree exchange diagrams as we considered only the Hartree term
in the electronic self--energy. However, it also includes phonon--mediated scatterings as a consequence of the fact that $\Sigma^{TP_{e-n}}$ has been included in the diagrammatic expansion. 
As it will be clear in the Sec.~\ref{sec:HF}, the addition of more diagrams to the electronic self--energy corresponds to modify the 
equation of motion satisfied by the dielectric function.
}}
\label{fig:6}
\end{figure}

The first two diagrams resulting from the expansion of the diagrammatic fraction appearing in Fig.~(\ref{fig:5}.c) are shown in Fig.~\ref{fig:6}. 
The repeated closed loops represent the Time--Dependent Hartree\,(TDH) approximation for the response function. The corresponding screening of the interaction is known as
Random--Phase Approximation\,(RPA).
The RPA is the most elemental way to introduce screening in a system of correlated electrons.

Self--consistency dresses the \epi in the Hartree, in the tad--pole and in the DW diagrams. As it will be clear in the following,
the equation of motion for the corresponding screening function changes with the level of approximation used in the self--energy.
Moreover, when $\Sigma^{TP_{e-n}}\(1\)+\Sigma^{n-n}\(1\)\neq 0$, the screening is due to the total time--dependent Hartree dielectric function that
includes the lattice polarization contribution. This follows from the definition of the zig--zag interaction, Fig.~\eqref{fig:4}  and  Eq.~\eqref{eq:3.11a},
which induces (see Fig.~\eqref{fig:6}, for example) scatterings processes where an electron--hole pair is annihilated and a phonon propagator is created.

Even today, most of the calculations at the $GW$ level are performed by using the $G_0W_0$ non self--consistent version. 
However, this section shows that the DW diagram gets correctly screened only by solving the Dyson equation self--consistently. 

%%%%%%%%%%%%%%%%%%%%%%%%%%%%%%%%%%%%%%%%%%%%%%%%%%%%%%%%%%%%%%%%%%%%%%%%%%%%%%%%%%%%%%%%%%%%%%%%%%%%%%%%%%%%%%%%%%%%%%%%%%%%%%%%%%%%%%%%%%%%%%%%%%%%%
\subsection{The Fock and Fan self-energies}
\label{sec:HF}
%%%%%%%%%%%%%%%%%%%%%%%%%%%%%%%%%%%%%%%%%%%%%%%%%%%%%%%%%%%%%%%%%%%%%%%%%%%%%%%%%%%%%%%%%%%%%%%%%%%%%%%%%%%%%%%%%%%%%%%%%%%%%%%%%%%%%%%%%%%%%%%%%%%%%
The analysis of the previous paragraph has been restricted at the Hartree level to keep the notation as simple as possible. In this section, we extend the derivation to the Fock approximation
by showing how the screening of the second--order \ep\,interaction $\theta$ is modified by the inclusion of electronic exchange scatterings via the
Fock diagram. We will assume for simplicity that $\Sigma^{e-n}+\Sigma^{n-n}=0$ and show the changes that this approximation induces in the definition of the dielectric function.

\begin{figure}[H]
\begin{center}
\epsfig{figure=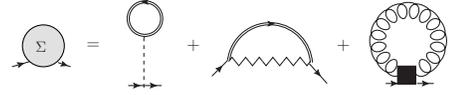,width=6cm}  %Diagrams/sigma_hf
\end{center}
\caption{\footnotesize{
The Dyson equation at the Fock level in the electron--electron and electron--phonon interaction. 
In addition to the Hartree and DW contributions we include the Fock diagram. Note that the zig--zag interaction lines
include the bare \ep\,interaction defined in  Fig.\eqref{fig:4}. Thus $\Sigma$ includes the DW and  the Fan diagrams with bare and 
un--dressed interactions.
}}
\label{fig:7}
\end{figure}

\begin{figure}[H]
\parbox[c]{8cm}{
\begin{center}
\epsfig{figure=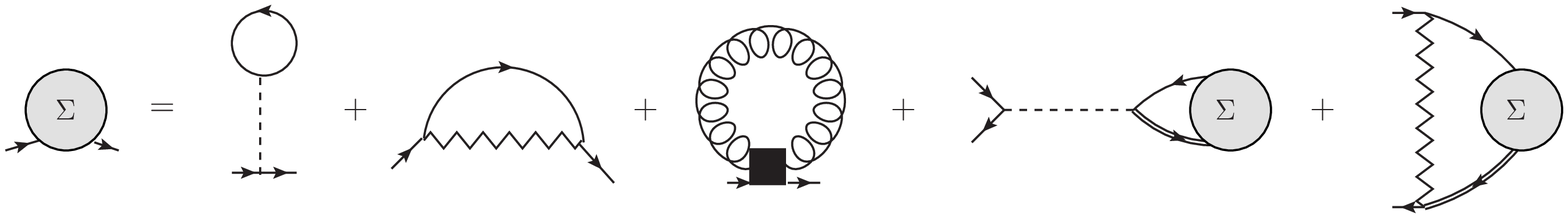,width=8cm}\\ %Diagrams/sigma_hf_1
(\ref{fig:8}.a)
\end{center}
}
\parbox[c]{8cm}{
\begin{center}
\epsfig{figure=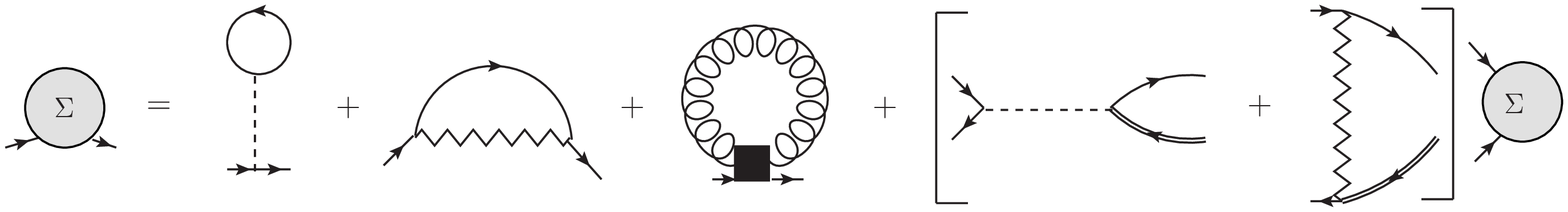,width=8cm}\\ %Diagrams/sigma_hf_2
(\ref{fig:8}.b)
\end{center}
}
\parbox[c]{8cm}{
\begin{center}
\epsfig{figure=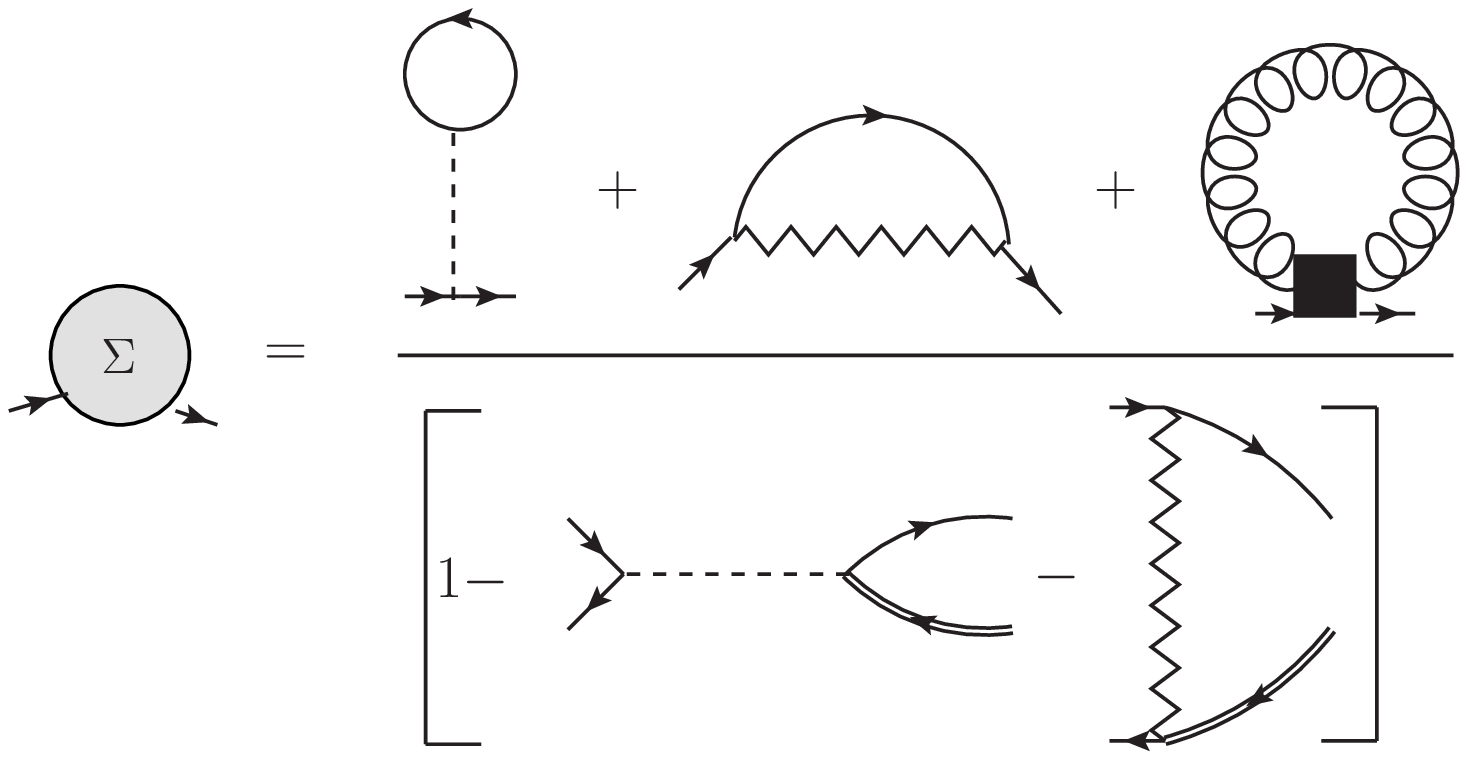,width=8cm}\\  %Diagrams/sigma_hf_3
(\ref{fig:8}.c)
\end{center}
}
\caption{\footnotesize{
Diagrammatic proof of the equivalence between self--consistency and screening when a Fock and a Fan diagrams are present (second and third in
the r.h.s. of Fig.\eqref{fig:7}).  In the frame (b) a portion of the equation is isolated and enclosed in square brackets. By 
defining formally the diagrammatic fraction (see text) the equation is inverted and the diagram in the square bracket goes
in the denonimator of frame (c). The perturbative expansion of this fraction leads to the screening of Hartree term and of
the second--order bare \ep\,interaction $\gt$. Some of the diagrams that follows from the expansion are shown 
in Fig.\eqref{fig:9}.
}}
\label{fig:8}
\end{figure}

The procedure to disentangle the self--consistency from the electron propagator appearing in the Fock operator can be done entirely using Feynman diagrams. 
We start by using the Dyson equation to rewrite the dressed (thick line) electronic propagators in Fig.~\eqref{fig:7} in terms of the bare electronic propagators (thin line) and
of the self--energy. The resulting diagrammatic expression for $\Sigma$ is showed in Fig.(\ref{fig:8}.a).

\begin{figure}[H]
\parbox[c]{8cm}{
\begin{center}
\epsfig{figure=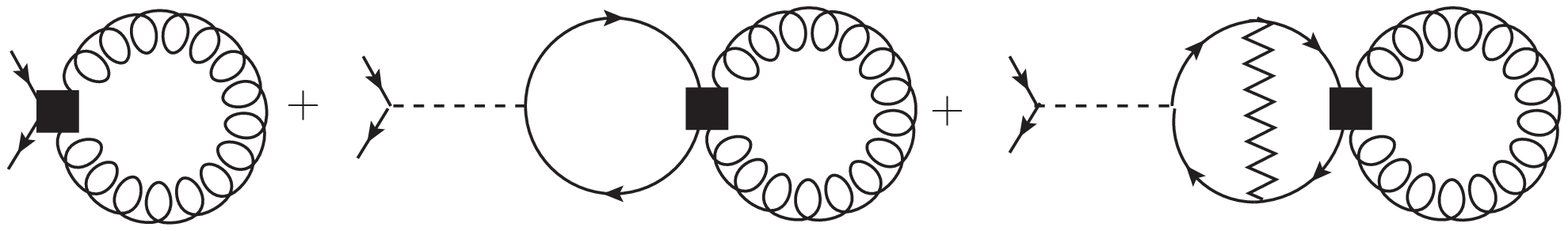,width=8cm}\\ %Diagrams/DW_tdhf.eps
(\ref{fig:9}.a)
\end{center}
}
\parbox[c]{8cm}{
\begin{center}
\epsfig{figure=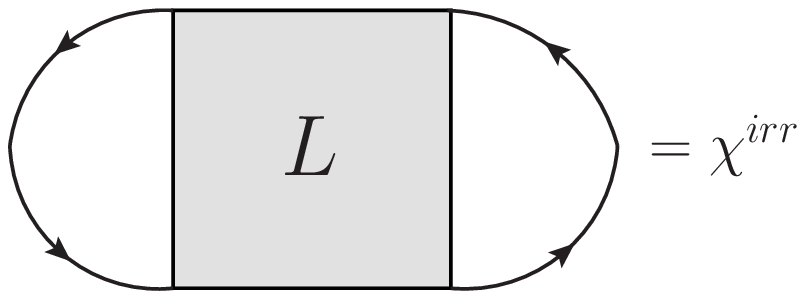,width=4cm}\\ %Diagrams/BSE_tdHF_1.eps
(\ref{fig:9}.b)
\end{center}
}
\parbox[c]{8cm}{
\begin{center}
\epsfig{figure=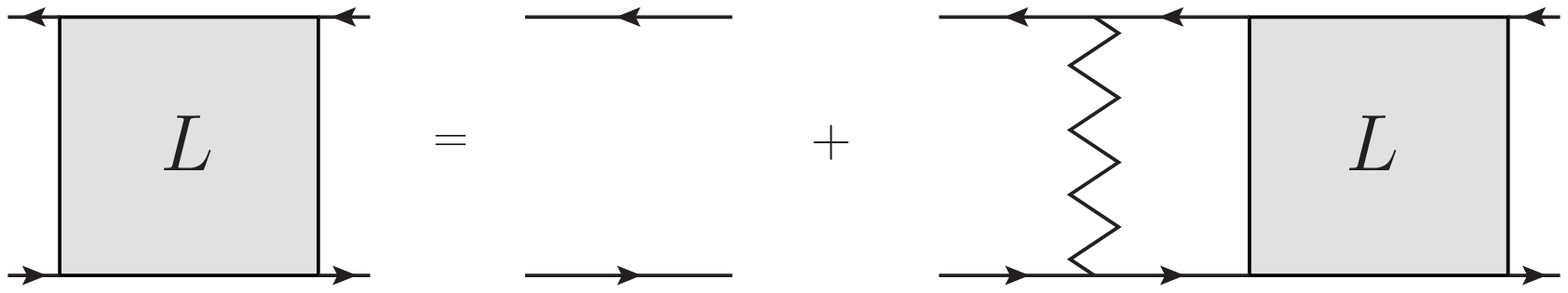,width=8cm}\\ %Diagrams/BSE_tdHF_2.eps
(\ref{fig:9}.c)
\end{center}
}
\caption{\footnotesize{
The series $(a)$ corresponds to the first three diagrams contributing to the screening of the DW term. Similar series of diagrams contributing to the self--energy can be obtained by replacing the DW  bare
self--energy with the Hartree and Fock self--energies. As we see besides the usual time--dependent Hartree contribution to the polarization
function (second diagram) there is a irreducible diagram where the electron and hole interact via the total screened interaction defined in
Eq.~\eqref{eq:3.11a}. The definition of the irreducible time--dependent Fock response function is given in the diagram $(b)$ in terms of the four point function $L$. The final equation for
$\chi^{irr}$ follows, then, from the corresponding equation of motion for $L$ (diagram $(c)$).
}}
\label{fig:9}
\end{figure}

We can then work out diagram~(\ref{fig:8}.a) by isolating a non--local operator that is evidenced by the square brackets in the diagram~(\ref{fig:8}.b). 
The self--energy can be isolated to the left-hand side of the equation. Then, in the same spirit as for the
Hartree case, we can invert the equation. By following the same procedure used to go from diagram~(\ref{fig:5}.b) to~(\ref{fig:5}.c) we 
introduce a diagrammatic fraction represented by the square bracket in the diagram~(\ref{fig:8}.c). Formally speaking, this fraction must be
interpreted as follows. Let us consider the generic expression
$\frac{1}{\hat{1}-\hat{D}_1-\hat{D}_2}$, with  $\hat{D}_1$ and $\hat{D}_2$ two generic diagrams with $M$ open interaction lines (in the present case 
 $M=4$). Then we have that
\begin{align}
\frac{1}{\hat{1}-\hat{D}_1-\hat{D}_2}\equiv\hat{1}+\sum_n \( \hat{D}_1+\hat{D}_2 \)^n_{c},
\label{eq:3.19a}
\end{align}
where the $c$ sub--script means that, at each order, we consider the totally contracted products of $\hat{D}_1$ and $\hat{D}_2$ 
in such a way that the resulting diagram has, again, $M$ open interaction lines. 
We can illustrate this procedure by applying it to the DW term (last term in the numerator of diagram~\ref{fig:8}.c).
In this case, Eq.~\eqref{eq:3.19a} applied to the square bracket produces an infinite series of diagrams that are closed in the upper part by a fermion 
line contracted in the second--order bare interaction ($\blacksquare$). 
The first three diagrams of this series are shown in Fig.~(\ref{fig:9}.a).

The final result is
that, like in Sec.\ref{sec:DW}, the Debye--Waller diagram is screened by a dielectric function. However, there are two important differences with respect to the Hartree case.
First of all, after linearization of the Green's functions appearing in the right-hand side of the diagram (\ref{fig:8}.c) by using Eq.~\eqref{eq:3.10d}, we can define a different dielectric function than Eq.~\eqref{eq:3.9p}:
\begin{equation}
\gee^{tdhf}_{\qq}\(1,2\)\equiv  \gd\(1,2\) - \int\,d3  v_{\qq}\(1,3\) \chi^{irr}_{\qq}\(3,2\),
\label{eq:3.20}
\end{equation}
with $\chi^{irr}_{\qq}\(3,2\)$ the time--dependent Hartree--Fock irreducible response function. The equation that defines $\chi^{irr}$ is represented in diagrammatic form in Fig.~(\ref{fig:9}.b) and Fig.~(\ref{fig:9}.c). 

The corresponding definition of the screened second--order \epi is
\begin{multline}
\widetilde{\gt}_{\qq\gl,-\qq\gl}\(1\)=\\
\int d2  \[\limq\gee_{\qq}^{tdhf}\(1,2\)\]^{-1} \theta_{\qq\gl,-\qq\gl}\(\rr_2\).
\label{eq:3.20p}
\end{multline}

The first two orders of $\chi$ are represented by the two bubbles 
appearing in Fig.(\ref{fig:9}.a) while in Fig.~\eqref{fig:6} only the independent-particle bubbles appear (this is, indeed, the definition of the RPA). 
In this case the Fock and Fan diagrams induce a first order bubble with the interaction $W$ connecting the two fermion propagators. This diagram represents the contribution of the electron--hole attraction and, when summed to all orders,
it can explain and predict the formation of excitonic states\cite{Onida2002}. Such bound electron--hole states are commonly observed in the absorption spectrum of several materials~\cite{Onida2002}.
In this case the electron--hole attraction is both electron and phonon mediated~\cite{Marini2008}.

The second difference is the fact that in this derivation of the Hartree--Fock screening, we have assumed that the two tad-poles  $\Sigma^{e-n}\(1\)$ and $\Sigma^{n-n}\(1\)$ cancel each other.
Formally speaking this cancellation is never exact. If the 
contribution of these two terms is included then the dashed interaction in the denominator of the diagrammatic equation (\ref{fig:8}.c) would contain a \ep contribution and 
the whole definition of the screening function would be affected. This corresponds, for example, to the appearance of 
exchange diagrams mediated by phonons.

%%%%%%%%%%%%%%%%%%%%%%%%%%%%%%%%%%%%%%%%%%%%%%%%%%%%%%%%%%%%%%%%%%%%%%%%%%%%%%%%%%%%%%%%%%%%%%%%%%%%%%%%%%%%%%%%%%%%%%%%%%%%%%%%%%%%%%%%%%%%%%%%%%%%%
\subsection{Skeleton diagrams, GW approximation and self--consistency issues}
\label{sec:GW}
%%%%%%%%%%%%%%%%%%%%%%%%%%%%%%%%%%%%%%%%%%%%%%%%%%%%%%%%%%%%%%%%%%%%%%%%%%%%%%%%%%%%%%%%%%%%%%%%%%%%%%%%%%%%%%%%%%%%%%%%%%%%%%%%%%%%%%%%%%%%%%%%%%%%%
In Sec.~\ref{sec:lowest_order} and Sec.~\ref{sec:HF} we have noticed that, even if we assume that tad--poles diagrams cancel each other, i.e. Eq.~\eqref{eq:3.10b} is
satisfied, the screening of the first ($\xi$) and second ($\theta$)
\ep interaction potentials induced by self--consistency depends on the kind of approximation used for the electronic self--energy.

\begin{figure}[H]
\epsfig{figure=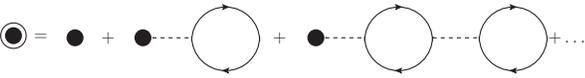,width=8cm} %Diagrams/GW_2.eps
\caption{\footnotesize{
Diagramatic representation of the dressed electron--phonon vertex within the $GW$ approximation for the self--energy. In this case the diagrams are bare\,(skeleton) and sum into an RPA dielectric screening
of the ionic potential. In this case, even in the case where the tad--pole diagrams are non zero, the dielectric screening is purely electronic. The standard additional approximation is to
consider a static screening.
}}
\label{fig:11}
\end{figure}

The situation for the other family of diagrams that must be considered at each order of the perturbative expansion is different. Indeed, if we consider
 skeleton (bare) diagrams we have that the $\xi$ function is renormalized by the purely electronic dielectric function, as explained for example 
in Ref.~\onlinecite{mattuck} (via the diagrammatic method) and in Ref.~\onlinecite{Leeuwen2004a} (via the equation of motion approach): 
\begin{equation}
\widetilde{\xi}_{\qq\gl}\(1\)=\int d2 \[\left.\gee_{\qq}\(1,2\)\right|_{el}\]^{-1} \xi_{\qq\gl}\(\rr_2\).
\label{eq:3.30}
\end{equation}
In this case $\[\left.\gee_{\qq}\right|_{el}\]^{-1}$ is the electronic dielectric function whose irreducible response function part follows directly by contracting the
vertex function associated with the self--energy. In the case of the well--known $GW$ approximation, the dielectric function
is calculated within the RPA. The purely electronic expression can be obtained
from Eq.~\eqref{eq:3.9p} when the phonon--mediated exchange
contribution is neglected and corresponds to approximate $W_{\qq\gl}\(1,2\)$ by $v_{\qq}\(\rr_1,\rr_2\)$. 
\begin{figure}[H]
\epsfig{figure=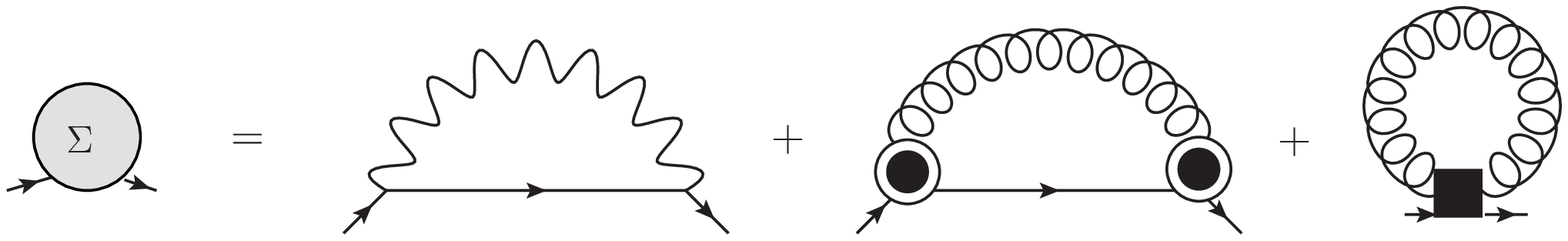,width=8cm} %Diagrams/GW_1.eps
\caption{\footnotesize{
The Dyson equation at the GW level in the electron--electron and electron--phonon interaction. The electron--phonon diagram is known as Fan self--energy and its vertex (represented by the
circled dot) represents a dressed electron--phonon interaction (see Eq.~\eqref{eq:3.30}). 
The wiggled line is a dressed electron--electron interaction (see Eq.~\eqref{eq:3.31}). The
most important aspect of this diagram is that, as long as only skeleton diagrams are included, the second--order \ep\,interaction, and consequently the 
DW diagram, is not screened.
}}
\label{fig:10}
\end{figure}
The non self--consistent self--energy is showed in Fig.~\eqref{fig:10}. The circled dot symbol (\epsfig{figure=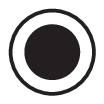,width=0.5cm}) represents
the dressed interaction defined in Eq.~\eqref{eq:3.30} and, diagrammatically, in  Fig.~\eqref{fig:11}. 
Similarly, the wiggled line is the screened electron--electron interaction
\begin{equation}
\widetilde{W}\(1,2\)=\int d3 \[\left.\gee\(1,3\)\right|_{el}\]^{-1} v\(2,3\).
\label{eq:3.31}
\end{equation}
All the equations and definitions connected to the inclusion of skeleton diagrams are well--known in the literature. The original aspect outlined by the derivations presented in the previous sections is that, while
the first order \ep\,interaction $\xi$ appearing in the Fan diagram is screened by skeleton diagrams, the second--order $\gt$ is screened by
self--consistency. 

This deep difference in the procedure that defines the kind of screening of the \epi is reflected in the different equation that is satisfied by $\gee$ in, for example,
Eq.~\eqref{eq:3.31}, Eq.~\eqref{eq:3.20p} and Eq.~\eqref{eq:3.12}. Depending on the choice of the self--energy, we have phonon mediated exchange and/or direct scatterings and electron--hole attraction diagrams. 
We will see in the following that yet another family of dielectric functions is used within the DFPT approach. The
physical interpretation of these different definitions is discussed in Sec.~\ref{sec:screening}.

%%%%%%%%%%%%%%%%%%%%%%%%%%%%%%%%%%%%%%%%%%%%%%%%%%%%%%%%%%%%%%%%%%%%%%%%%%%%%%%%%%%%%%%%%%%%%%%%%%%%%%%%%%%%%%%%%%%%%%%%%%%%%%%%%%%%%%%%%%%%%%%%%%%%%
\section{The Density--Functional--Theory approach to the \ep problem}
\label{sec:the_dfpt}
%%%%%%%%%%%%%%%%%%%%%%%%%%%%%%%%%%%%%%%%%%%%%%%%%%%%%%%%%%%%%%%%%%%%%%%%%%%%%%%%%%%%%%%%%%%%%%%%%%%%%%%%%%%%%%%%%%%%%%%%%%%%%%%%%%%%%%%%%%%%%%%%%%%%%
In the previous section we have analyzed the kind of diagrams induced by the electron--electron interaction in the dressing of the 
electron--nucleus interaction terms. We have disclosed the key role played by self--consistency and the different level of approximation that arises 
from the perturbative expansion. 

In order to link the DFPT and MBPT approaches we 
start with a short review of the purely DFT--based approach to the electron--phonon coupling.

DFT is a self--consistent theory,
and DFPT is its extension to take into account, self--consistently, the effect of static perturbations (like nuclear displacements). In this
case, DPFT provides an exact description of phonons within the limits of a static and adiabatic approach. The phonon frequencies in DFPT are always real and no
phonon dissipation process is included.
If we introduce a total (electron--electron plus electron--nucleus) potential
\begin{equation}
\widehat{V}_{scf}\(\RR,\rr\)=  \widehat{V}_{Hxc}\[\gr\]\(\rr\)-\sum_{ls}\frac{Z_s}{|\hat{\rr}-\hat{\RR}_{ls}|},
\label{eq:4.4}
\end{equation}
where the functional dependence of the electronic density on the nuclear positions introduce a direct (via $W_{e-n}$) and 
indirect (via $V_{Hxc}$) dependence on $\(\RR\)$ in $V_{scf}$. In DFPT this complicated dependence links the calculation
of $\partial_{\RR_{ls}} \widehat{V}_{scf}\(\RR,\rr\)$ to the solution of a self--consistent set 
of equations.

By applying the same procedure used to derive Eq.~\eqref{eq:1.4}, a formal Taylor expansion of $H_{KS}$ can be obtained. 
However, if the dependence of the density on the nuclear positions is not taken into account,
all terms in the Taylor expansion are bare. In DFT (or DFPT), screening builds--up because of the
$\widehat{V}_{scf}$ indirect dependence on $\(\RR\)$. We introduce
\begin{equation}
\xi^{DFPT}_{\gql}\(\rr\)=\partial_{\(\qq\gl\)} V_{scf}\(\RR,\rr\),
\label{eq:4.5}
\end{equation}
and
\begin{equation}
\theta^{DFPT}_{\gql,-\qq\gl}\(\rr\)=\partial_{\(\qq\gl\),\(-\qq\gl\)} V_{scf}\(\RR,\rr\).
\end{equation}
The expression for $\xi^{DFPT}_{\gql}\(\rr\)$ can be written in terms 
of $\partial_{\RR_s} V_{scf}\(\RR,\rr\)$ by following the same procedure outlined in appendix~\ref{appB}. The screening 
of $\xi$ within a pure DFPT scheme follows from the fact that
\begin{multline}
\partial_{\RR_{ls}} V_{scf}\(\RR,\rr\)=\partial_{\RR_{ls}} W_{e-n}\(\RR,\rr\)\\
+\int\,d\rr'\frac{\gd V_{Hxc}\[\gr\]\(\rr\)}{\gd\gr\(\rr'\)}\partial_{\RR_{ls}}\gd\gr\(\rr'\).
\label{eq:4.6}
\end{multline}
In order to evaluate Eq.~\eqref{eq:4.6} and create a link with the MB approach, we notice that DFPT is based on the linear response
regime~\cite{baroni2001,Gonze1995} where:
\begin{equation}
\partial_{\RR_{ls}}\gr\(\rr\)=\int\,d\rr' \chi_{DFT}\(\rr,\rr'\) \partial_{\RR_{ls}} W_{e-n}\(\RR,\rr'\),
\label{eq:4.7}
\end{equation}
where the DFT polarizability $\chi_{DFT}=\frac{\partial\rho}{\partial V_{ext}}$ is solution of the following Dyson equation
\begin{multline}
\chi_{DFT}\(\rr,\rr'\)=\chi^{\(0\)}_{KS}\(\rr,\rr'\)\\
+\int\,d\rr^{''}d\overline{\rr}\chi^{\(0\)}_{KS}\(\rr,\overline{\rr}\)f_{Hxc}(\overline{\rr},\rr^{''})\chi_{DFT}(\rr^{''},\rr'),
\label{eq:4.9a}
\end{multline}
with $\chi^{\(0\)}_{KS}$ the independent particle KS response function.

From the definition of the Hartree and xc potential it follows that
\begin{align}
f_{Hxc}\(\rr,\rr'\)&\equiv\frac{\gd V_{Hxc}\[\gr\]\(\rr\)}{\gd\gr\(\rr'\)}\\
&=v\(\rr-\rr'\)+f_{xc}\(\rr,\rr'\),
\label{eq:4.8}
\end{align}
which, finally, yields the well-known expression for the derivative of the total DFT self--consistent potential
\begin{multline}
\partial_{\RR_{ls}}
V_{scf}\(\RR,\rr\)=\int\,d\rr'\Bigl\{\gd\(\rr-\rr'\)\\
+\int\,d\rr^{''}f_{Hxc}\(\rr,\rr^{''}\)\chi_{DFT}\(\rr^{''},\rr'\)\Bigr\}\\
\partial_{\RR_{ls}}W_{e-n}\(\RR,\rr'\).
\label{eq:4.9}
\end{multline}
If we now introduce the DFT dielectric function
\begin{multline}
\[\gee^{DFT}\(\rr,\rr'\)\]^{-1}\equiv\gd\(\rr-\rr'\)\\
+\int\,d\rr^{''}f_{Hxc}\(\rr,\rr^{''}\)\chi_{DFT}\(\rr^{''},\rr'\),
\label{eq:4.10}
\end{multline}
we have, finally, that
\begin{multline}
\partial_{\RR_{ls}}
V_{scf}\(\RR,\rr\)=\int\,d\rr'\[\gee^{DFT}\(\rr,\rr'\)\]^{-1}\\
\times \partial_{\RR_{ls}} W_{e-n}\(\RR,\rr'\).
\label{eq:4.11}
\end{multline}
Similarly, the second--order derivative of $\Vscf$ can be introduced~\cite{Gonze1995,baroni2001} as
\begin{multline}
\partial^2_{\RR_{ls}\RR_{l's'}}
V_{scf}\(\RR,\rr\)=\\
\partial_{\RR_{ls}}\Biggl\{ \int\,d\rr'\[\gee^{DFT}\(\rr,\rr'\)\]^{-1}
\partial_{\RR_{l's'}} W_{e-n}\(\RR,\rr'\)\Biggr\}.
\label{eq:4.12}
\end{multline}
Eqs.~\eqref{eq:4.11} and \eqref{eq:4.12} must be compared with Eqs.~\eqref{eq:3.12}, \eqref{eq:3.12.1} and \eqref{eq:3.12.2} in order to highlight the differences between the two formulations and potential similitudes. 

Eq.~\eqref{eq:4.11} can be written in the basis of phonon displacements $\(\qq,\gl\)$ as
\begin{equation}
\widetilde{\xi}^{DFPT}_{\qq\gl}\(\rr\)=\int\,d\rr' \[\gee^{DFT}_{\qq}\(\rr,\rr'\)\]^{-1} \xi^{DFPT}_{\qq\gl}\(\rr'\).
\label{eq:4.13}
\end{equation}  
This last equation can directly be compared with Eq.~\eqref{eq:3.12} and the following observations can be made
\begin{itemize}
\item[(a)] In DFPT the $\xi$ function (and more generally the electron--phonon interaction) is statically screened and it does not include the contribution from the lattice polarization. In the MBPT, instead,
the \epi is dynamically screened (i.e. the dielectric functions defined in Eq.~\eqref{eq:3.9p} and Eq.~\eqref{eq:3.20} are time dependent) and it includes
phonon mediated scatterings (see Eq.~\eqref{eq:3.12}). 
The static screening of DFPT reflects the fact that there are no retardation effects in the theory. Those effects are peculiar of the MBPT and, in some cases, can 
lead to important deviations from the static limit~\cite{cannuccia_thesis,cannuccia,Cannuccia2012} when included in the self--energy. 
\item[(b)] The DFT dielectric function $\gee^{DFT}$ defined in Eq.~\eqref{eq:4.10} is a test--electron dielectric function whereas in the MBPT the dielectric function that screens the bare
\epi is a test--charge function. The difference between those two functions is well described in Ref.~\onlinecite{Hybertsen1987} and~\onlinecite{Ghosez1997}. In the 
test--charge case, the dielectric function represents a response to an external particle, while in the test--electron case, the charge is itself an electron. We will discuss in more detail this
difference from a physical perspective in Sec.\ref{sec:screening}.
\item[(c)]  A peculiar consequence of the DFPT approach is the appearance of non--rigid nuclei contributions to the second--order derivative of the electron--nucleus interaction potential,
$\partial^2_{\RR_{ls}\RR_{l's'}} V_{scf}\(\RR,\rr\)$. This contribution does not appear in the Many--Body derivation carried on in the previous sections.
\end{itemize}

Such non--rigid nuclei (recently called in the literature non--rigid ion contribution\cite{SP_2014}) can be studied by applying the derivative with respect to a nucleus displacement on the right-hand side of Eq.~\eqref{eq:4.12} and by distinguishing a rigid--nuclei (RN) and a non--rigid nuclei (NRN) terms 
\begin{multline}
\partial^2_{\RR_{ls}\RR_{l's'}} V_{scf}\(\RR,\rr\)= 
\left.\[\partial^2_{\RR_{ls}\RR_{l's'}} V_{scf}\(\RR,\rr\)\]\right|_{RN}\\
+\left.\[\partial^2_{\RR_{ls}\RR_{l's'}} V_{scf}\(\RR,\rr\)\]\right|_{NRN},
\label{eq:4.14}
\end{multline}
with
\begin{multline}
\left.\[\partial^2_{\RR_{ls}\RR_{l's'}} V_{scf}\(\RR,\rr\)\]\right|_{RN}= \\
\int\,d\rr'  \[\gee^{DFT}\(\rr,\rr'\)\]^{-1} \partial^2_{\RR_{ls}\RR_{l's'}} W_{e-n}\(\RR,\rr'\),
\label{eq:4.15}
\end{multline}
and
\begin{multline}
\left.\[\partial^2_{\RR_{ls}\RR_{l's'}} V_{scf}\(\RR,\rr\)\]\right|_{NRN}= \\
\int\,d\rr' \partial_{\RR_{ls}} \[\gee^{DFT}\(\rr,\rr'\)\]^{-1}\partial_{\RR_{l's'}} W_{e-n}\(\RR,\rr'\).
\label{eq:4.15p}
\end{multline}
If we now rewrite both terms in the DFPT phonon $\(\qq,\gl\)$ basis and multiply by a $1/2$ pre factor we get
\begin{multline}
\widetilde{\gt}_{\qq\gl,\qq'\gl'}^{DFPT,RN}\(\rr\)\equiv \frac{1}{2}
  \int\,d\rr' \[\gee^{DFT}\(\rr,\rr'\)\]^{-1}  \\
  \times \partial^2_{\qq\gl,\qq'\gl'} V_{scf}\(\RR,\rr'\),
\label{eq:4.15s}
\end{multline}
and
\begin{multline}
\widetilde{\gt}_{\qq\gl,\qq'\gl'}^{DFPT,NRN}\(\rr\)\equiv\\
\frac{1}{2} \int\,d\rr' \widetilde{\xi}^{DFPT}_{\qq'\gl'}\(\rr'\) \partial_{\qq\gl} \[\gee^{DFT}\(\rr,\rr'\)\]^{-1}.
\label{eq:4.15t}
\end{multline}
From the local dependence on $\RR$ of the $V_{scf}$ it follows that, in the RN contribution, a $\gd_{\RR_{ls},\RR_{l's'}}$ appears. In the rigid--nuclei approximation
(also called the rigid-ion approximation) the $\widetilde{\gt}_{\qq\gl,\qq'\gl'}^{DFPT,NRN}$ is neglected.

The physical interpretation of the NRN contribution to the second--order derivative of the self--consistent potential is obscure in the DFPT derivation and
seems to be more a mathematical separation based on computational load of the calculations rather than a physically motivated choice (see Ref.~\onlinecite{SP_2014} for a detailed explanation).
We will discuss this term from a Many--Body perspective in Sec.~\ref{sec:nddw}.

%%%%%%%%%%%%%%%%%%%%%%%%%%%%%%%%%%%%%%%%%%%%%%%%%%%%%%%%%%%%%%%%%%%%%%%%%%%%%%%%%%%%%%%%%%%%%
\section{MBPT starting from Density--Functional and Density--Functional Perturbation Theory}
\label{sec:merge}
%%%%%%%%%%%%%%%%%%%%%%%%%%%%%%%%%%%%%%%%%%%%%%%%%%%%%%%%%%%%%%%%%%%%%%%%%%%%%%%%%%%%%%%%%%%%%
The main difficulty in performing a diagrammatic expansion on top of DFT is that it is not possible to write the initial Hamiltonian in terms of
dressed interactions. This means that, even within DFT, the Taylor expansion of $\widehat{H}\(\RR\)$ is still given by Eq.~\eqref{eq:2.3a} with the only
difference that the electron--electron interaction is replaced by the mean--field xc--potential. At this point DFPT and MBPT follow two different routes in order to describe the dressing
of the interaction and the definition of the electronic self--energies.

To obtain the same screening as DFT+DFPT from a purely many--body point of view, we start by writing again the total Hamiltonian as a function of the KS one
\begin{multline}
\widehat{H}\(\RR\)= \widehat{H}_{KS}\(\overline{\RR}\)+\widehat{H}_{n}\(\overline{\RR}\)+
\Delta\whW^{ref}_{n-n}\(\RR\) \\
+\Delta \widehat{H}_{remaining}\(\RR\).
\label{eq:5.1}
\end{multline}
We can determine the value of $\Delta \widehat{H}_{remaining}\(\RR\)$ from Eqs.~\eqref{eq:1.7}, \eqref{eq:ip.2} and \eqref{delta_Hamil}
\begin{multline}
\Delta \widehat{H}_{remaining}\(\RR\) = \Delta \whW_{n-n}\(\RR\) - \Delta\whW^{ref}_{n-n}\(\RR\)  \\
  + \Delta\whW_{e-n}\(\RR\)+\whW_{e-e}-\widehat{V}_{Hxc}\[\overline{\gr}\],
\label{eq:5.2}
\end{multline}
with $\overline{\gr}$ the equilibrium DFT density. 

Now the full MBPT machinery described in the previous sections can be applied to Eq.~\eqref{eq:5.2}, leading to the screening of the \ep interactions. However, our aim is to create a link with
the DFPT definitions of Eq.~\eqref{eq:4.11} and \eqref{eq:4.12}. The main differences between the MBPT and DFPT approach
are listed in Table~\ref{tab:1} and explained below.

\begin{table*}
\caption{Schematic representation of the different treatment within MBPT and DFPT of the most important aspects of the electron--nucleus interaction.\label{tab:1}}
\begin{tabularx}{\textwidth}{ |X|X|X|X| }
\hline
& 
MBPT& 
DFPT \\\hline
Tad--Pole diagrams sum, $\Sigma^{e-n}+\Sigma^{n-n}$, Fig.~(\ref{fig:2}.b).
&Vanishing only when the atomic positions are evaluated consistently with the level of correlation included in the self--energy. 
&Vanishing. \\\hline
Screening of the first order interaction, $\xi\(\rr\)$.
&Time--dependent and described by a test--charge dielectric function consistent with the self--energy. Induced by skeleton diagrams. 
&Screened by the static test--electron DFT dielectric function. Introduced via self--consistency. \\\hline
Screening of the second order interaction, $\gt\(\rr\)$. 
&Time--dependent and described by a test--charge dielectric function consistent with the self--energy. Induced by self--consistency diagrams.
&Screened by the static test--electron DFT dielectric function. Introduced via self--consistency. \\\hline
Non--rigid nuclei contribution to the second order interaction, $\gt\(\rr\)$. 
&Appears in the static limit from the diagrams that describe the dressing of the electronic density caused by the e--n interaction. It is, again, caused 
by self--consistent diagrams.
&Present and static. Induced by the implicit dependence of the screening on the atomic positions.  \\
\hline
\end{tabularx}
\end{table*}

%%%%%%%%%%%%%%%%%%%%%%%%%%%%%%%%%%%%%%%%%%%%%%%%%%%%%%%%%%%%%%%%%%%%%%%%%%%%%%%%%%%%%%%%%%%%%
\subsection{Tad--Poles}
%%%%%%%%%%%%%%%%%%%%%%%%%%%%%%%%%%%%%%%%%%%%%%%%%%%%%%%%%%%%%%%%%%%%%%%%%%%%%%%%%%%%%%%%%%%%%
By definition the DFT density is the exact one and it satisfies Eq.~\eqref{eq:3.10}. Thus, bearing in mind the linearization procedure outlined at the end of Sec.~\ref{sec:tad_pole}, we can affirm
that within DFPT, $\Sigma^{e-n}\(1\)+\Sigma^{n-n}\(1\)=0$.  As a consequence the dielectric function that screens the \epi does not include
phonon-mediated exchange scatterings.

%%%%%%%%%%%%%%%%%%%%%%%%%%%%%%%%%%%%%%%%%%%%%%%%%%%%%%%%%%%%%%%%%%%%%%%%%%%%%%%%%%%%%%%%%%%%%
\subsection{Screening of the second--order electron--nucleus interaction}
\label{sec:screening}
%%%%%%%%%%%%%%%%%%%%%%%%%%%%%%%%%%%%%%%%%%%%%%%%%%%%%%%%%%%%%%%%%%%%%%%%%%%%%%%%%%%%%%%%%%%%%
In order to discuss how screening builds up, we start from the lowest order self--energy in the electron--phonon scattering (see Fig.(\ref{fig:2}.b)).
Again, we start from the bare Hamiltonian but this time we use the KS system as reference.
The Dyson equation that follows from
the Hamiltonian of Eq.~\eqref{eq:5.1} is
\begin{multline}
 G\(1,2\)=G^{KS}\(1,2\)+\int\,d3 G^{KS}\(1,3\)\\
 \times \Bigl\{\Sigma^{H}\(3\)-V_{Hxc}\[\overline{\gr}\]\(3\)\Bigr\}G\(3,2\).
\label{eq:5.2p}
\end{multline}
Moreover, since the Hartree part ($V_H$) also appears in $V_{Hxc}$, we have
\begin{multline}
\Sigma^{H}\[\gr\]\(1\)-V_{H}\[\overline{\gr}\]\(1\)= \\
\int\,d2 v\(1,2\)\Bigl[\gr\(2\)-\overline{\gr}\(2\)\Bigr].
\label{eq:5.3}
\end{multline}
In the case of a local self--energy, the difference of densities can be rewritten in terms of the self--energy by
using Eq.~\eqref{eq:3.7b}
\begin{equation}
\gr\(1\)-\overline{\gr}\(1\)=\int\,d2 G^{KS}\(1,2\)\Sigma\(2\) G\(2,1\).
\label{eq:5.4}
\end{equation}
Following Eq.~\eqref{eq:3.9p}, we obtain within RPA and after linearization of the Green's function ($G\approx G^{KS}$) that
\begin{equation}
\gee^{KS,RPA}\(1,2\)\equiv  \gd\(1,2\) 
- \int\,d3  v\(1,3\) \chi^{KS}\(3,2\),
\label{eq:5.4p}
\end{equation}
with
\begin{equation}
\chi^{KS}\(1,2\)\equiv G^{KS}\(1,2\) G^{KS}\(2,1\).
\label{eq:5.5}
\end{equation}
The final expression of the second--order \epi thus becomes
\begin{multline}
\widetilde{\gt}_{\qq\gl,-\qq\gl}\(1\)=\int d2  \[\limq \gee_{\qq}^{KS,RPA}\(1,2\)\]^{-1}\\
\times \theta_{\qq\gl,-\qq\gl}\(\rr_2\)\gd\(t_2\),
\label{eq:5.5p}
\end{multline}
with the dielectric function defined by
\begin{multline}
\[\gee^{KS,RPA}\(1,2\)\]^{-1}\equiv  \gd\(1,2\) \\
+ \int\,d3  v\(1,3\) \chi^{KS,RPA}\(3,2\),
\label{eq:5.5s}
\end{multline}
and $\chi^{KS,RPA}$ solution of Eq.~\eqref{eq:4.9a} with $f_{Hxc}\(\rr,\rr'\)=v\(\rr,\rr'\)$.

We notice that Eq.~\eqref{eq:5.5s} defines a test--charge dielectric function while in DFPT, Eq.~\eqref{eq:4.10}, a test--electron dielectric function appears.  

This difference of definition is a straightforward consequence of the fact that the
definition of the dielectric function is linked to the distinction between the classical and the quantum parts of the induced potential.
The classical part satisfies a Poisson equation whose solution is the Hartree potential. The quantum part is treated in different ways in
MBPT and DFPT. In the MBPT, the quantum induced field is represented by the change of the 
correlation self--energy due to a test charge. This is described by the vertex function\cite{Strinati1988} that can also be used to rewrite the exact self--energy in a closed form.
In DFT, instead, electronic correlations are included in a mean--field manner by means of the exchange--correlation $V_{xc}$ potential. It follows that,
the variation of $V_{xc}$ mimics the variation of the self--energy and, thus, represent the quantum part of the induced potential.
Therefore, the difference between a test--electron and a test--charge is that the test--electron includes the total variation of the {\em total} potential, 
including $V_{xc}$. This contribution leads to the $f_{xc}$ term appearing in the right-hand side of Eq.~\eqref{eq:4.10} and marks the difference between the
MBPT and DFPT screening.

As additional proof, we can notice that if the electronic self--energy is approximated with a DFT exchange--correlation potential, then 
the many--body vertex turns the test--particle into a test--electron dielectric function that is
consistent with the DFPT definition. This means that $f_{xc}$ is taking into account, in a mean--field manner, the effect of the MBPT vertex function.

At this point, we can conclude by observing that if  local or semi--local approximations for $V_{xc}$ are used, the difference between 
a test--charge and a test--electron dielectric function can be safely neglected. Indeed any local or semi--local expression for $V_{xc}$
is regular in the short--distance limit whereas $v\(\rr,\rr'\)$ diverges. This means that, when $\qq\rightarrow {\bf 0}$, it follows that
$f_{Hxc}\(\rr,\rr'\)\approx v\(\rr,\rr'\)$.

More elaborate expressions for $V_{xc}$ that also include proper short--distance spatial corrections exist. 
In this case, a more accurate analysis of the effect of the Many--Body vertex function on the screening of the 
\epi becomes essential to draw a conclusive parallel between MBPT and DFPT. However, this goes beyond the scope of the present work.

Therefore, as far as local or semi--local approximations for $V_{xc}$ are used, we can approximate the DW screened interaction with the one evaluated at the DFPT level. 

%%%%%%%%%%%%%%%%%%%%%%%%%%%%%%%%%%%%%%%%%%%%%%%%%%%%%%%%%%%%%%%%%%%%%%%%%%%%%%%%%%%%%%%%%%%%%
\subsection{The non--rigid nuclei contribution to the Debye--Waller self--energy from a diagrammatic perspective}
\label{sec:nddw}
%%%%%%%%%%%%%%%%%%%%%%%%%%%%%%%%%%%%%%%%%%%%%%%%%%%%%%%%%%%%%%%%%%%%%%%%%%%%%%%%%%%%%%%%%%%%%
One of the most undeniable difference between the DFT and MBPT scheme is the absence of a diagrammatic explanation for the non--rigid nuclei contribution (Eq.~\eqref{eq:4.15t}) to the
second--order derivative of the nuclear potential. This terms has been shown to be quite important in low--dimensional systems~\cite{Gonze2011} but, at first sight, it does not appear in
the many--body theory of the \epip\, Indeed, in Eq.~\eqref{eq:3.12.1} and Eq.~\eqref{eq:3.20p}, the derivatives with respect to the atomic positions act only on the local bare ionic 
potential.

\begin{figure}[H]
\begin{center}
\parbox[c]{8cm}{
\begin{center}
\epsfig{figure=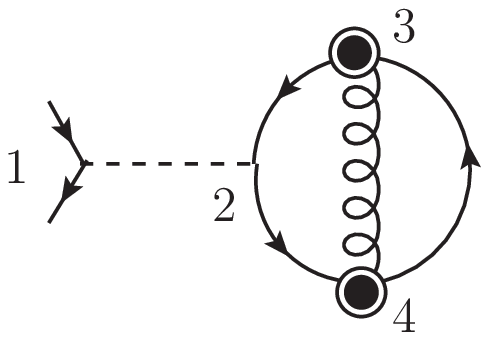,width=3cm}\\ %Diagrams/NDDW_1.eps
(\ref{fig:12}.a)
\end{center}
}\\
\parbox[c]{8cm}{
\begin{center}
\epsfig{figure=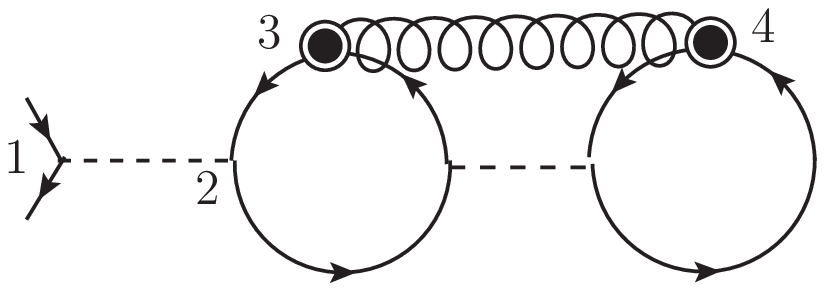,width=4cm}\\ %Diagrams/NDDW_2.eps
(\ref{fig:12}.b)
\end{center}
}\\
\parbox[c]{8cm}{
\begin{center}
\epsfig{figure=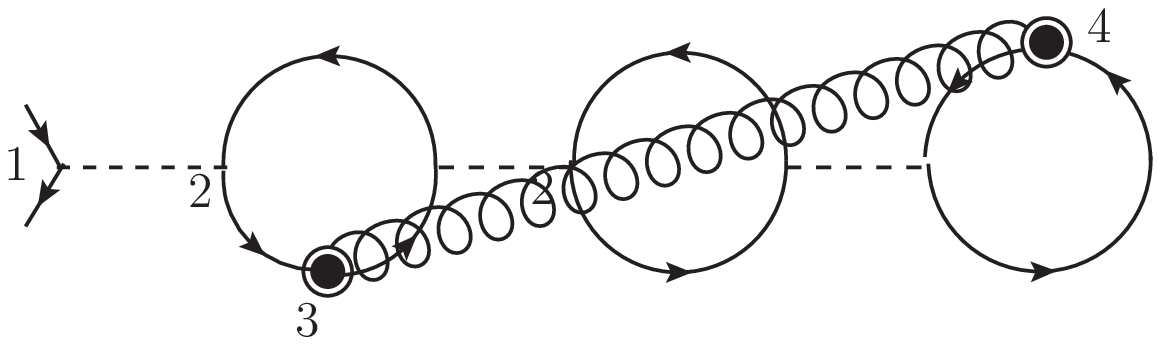,width=6cm}\\ %Diagrams/NDDW_3.eps
(\ref{fig:12}.c)
\end{center}
}\\
\end{center}
\caption{\footnotesize{First three  diagrams that dress the Hartree self--energy due to the change of the electronic density induced by the
\epip\, The sum of all diagrams of this kind (see Eq.~\eqref{eq:4.20} for a more formal definition) reduce, in the static and adiabatic limit, to the 
NRN contribution to the Debye--Waller self--energy as discussed in the text.}}
\label{fig:12}
\end{figure}

In order to find a diagrammatic perspective of $\widetilde{\gt}_{\qq\gl,\qq'\gl'}^{DFPT,NRN}$ we follow again the path of performing a diagrammatic expansion 
of the bare, un--dressed Hamiltonian and, at the end, draw links with DFT. In the present case we need to
to consider a new series of diagrams describing the change
of the electronic density induced by the electron--phonon interaction. Three examples of this series of diagrams are showed
in Fig.~\eqref{fig:12}. We can see that those diagrams are of the same order of magnitude as the lowest
order self--energies and there is \textit{a priori} no reason to neglect them.

The dressing of the Hartree self--energy can occur in two ways: the first is by dressing the internal electronic
propagators (an example is the Diagram\,(\ref{fig:12}.a)). The second way is to dress two tad--poles with phonon scatterings like in Diagrams\,(\ref{fig:12}.b)
and (\ref{fig:12}.c). All the diagrams of this kind can be written as
\begin{equation}
\gd\bgS^{H}\(\rr_1\)=  \int\,d\rr_2 v\(\rr_1-\rr_2\) \gd{\bf \gr}\(\rr_2\),
\label{eq:4.20}
\end{equation}
with 
\begin{align}
\gd\gr\(\rr_2\)&=\[\gr\(\rr_2\)-\overline{\gr}\(\rr_2\)\]=\sum_{\pp} \gd{\bf \gr}_{\pp}\(\rr_2\) \\
&=\frac{1}{N}\sum_{n\pp} \phi_{n\pp}\(\rr_2\) \phi^{*}_{n'\pp}\(\rr_2\)\gd\gr_{nn'\pp}\(\rr_2\).
\label{eq:4.20p}
\end{align}
In the case of  the diagram\,(\ref{fig:12}.a) we have that
\begin{multline}
\gd{\bf \gr}_{\pp}\(2\)=
\sum_{\qq,\gl} \int\,d34 \[\GG_{\pp}^{\(0\)}\(2,3\) \widetilde{\xi}_{\qq\gl}\(3\)  \capo\times  \GG^{\(0\)}_{\pp-\qq}\(3,4\)\] 
\widetilde{\xi}^{*}_{\qq\gl}\(4\)  \GG_{\pp}^{\(0\)}\(4,2\) D_{\qq\gl}^{\(0\)}\(t_3-t_4\),
\label{eq:4.22}
\end{multline}
where $\widetilde{\xi}$ is a first-order interaction that we assume to be screened by the very same skeleton diagrams that we
described in Sec.\ref{sec:GW}.

In order to introduce a simple and clear interpretation of the contribution due to the $\gd\Sigma^{H}$ diagrams, we take the static and adiabatic limit 
of the atomic displacements. This approach will also simplify the connection with the corresponding quantity evaluated with the
DFPT scheme. 

\begin{figure}[H]
$\partial_{\qq\gl}$
\parbox[c]{1cm}{
\epsfig{figure=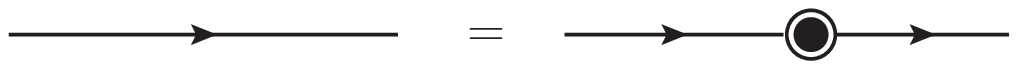,width=7.5cm} %Diagrams/partial_R_G.eps
}
\caption{\footnotesize{Diagrammatic transposition of the derivative of a Green's function with respect to a specific displacement written in the phonon basis (see text, in particular Eq.~\eqref{eq:4.19}). The derivative 
splits the Green's function with the insertion of an \ep\, first--order interaction.}}
\label{fig:13}
\end{figure}

We start by taking the static limit of $D_{\qq\gl}^{\(0\)}\(t_3-t_4\)\approx -\gd\(t_3-t_4\) \[2 n\(\go_{\qq\gl}\) +1\]$ and of $\widetilde{\xi}\(3\)\approx \widetilde{\xi}\(\rr_3\)\gd\(t_3\)$. 
Moreover, in this limit, we can treat the 
atomic displacements as classical and static variables in order to approximate
$\Delta \widehat{H}\(\RR\)$ 
\begin{equation}
\Delta{\widehat{H}\(\RR\)}\approx \sum_{\qq\gl,i} \widetilde{\xi}_{\qq\gl}\(\rr_i\) u_{\qq\gl},
\label{eq:4.16}
\end{equation}
with $u_{\qq\gl}$ the phonon displacements defined as
\begin{equation}
u_{\qq\gl}=\sum_{ls\ga} \(2 N M_s \go_{\qq\gl}\)^{-1} \Delta R_{ls\ga} \eta_{\ga}\(\qq\gl|s\).
\label{eq:4.16p}
\end{equation}
As we are interested in a specific series of diagrams, we can disregard higher-order corrections to $\Delta H\(\RR\)$.
The Eq.~\eqref{eq:4.16} allows to formally define
 the derivative of the Green's function, $\overline{\partial_{\qq\gl} G\(1,2\)}$ evaluated at the equilibrium nuclear positions.  Indeed, from Eq.~\eqref{eq:4.16}, it follows that
\begin{equation}
\[G\(1,2\)\]^{-1}=\[G^{\(0\)}\(1,2\)\]^{-1}-\Delta H\(\RR\).
\label{eq:4.17}
\end{equation}
and
\begin{multline}
\overline{\partial_{\qq\gl}\[G\(1,2\)\]^{-1}}= \\
- \partial_{\qq\gl} \Delta H\(\RR\) = - \widetilde{\xi}_{\qq\gl}\(\rr_1\)\gd\(1,2\),
\label{eq:4.19p}
\end{multline}
which, finally, using the identity
\begin{equation}
\gd\(1,2\)=\int\,d3 \[G\(1,3\)\]^{-1} G\(3,2\),
\label{eq:4.18}
\end{equation}
yields the desired definition
\begin{equation}
\overline{\partial_{\qq\gl}G\(1,2\)}=\int\,d3 G^{\(0\)}\(1,3\) \widetilde{\xi}_{\qq\gl}\(\rr_3\) G^{\(0\)}\(3,2\).
\label{eq:4.19}
\end{equation}
Eq.~\eqref{eq:4.19} is diagrammatically represented in Fig.~\eqref{fig:13}. The derivative of the Green's function splits the projector in two with the insertion
of a dressed \ep first--order interaction (\epsfig{figure=inline_figure.eps,width=0.5cm}).

By using Eq.~\eqref{eq:4.19} we notice that the first--order interactions appearing in the $\gr\(\rr_2\)-\overline{\gr}\(\rr_2\)$ can be interpreted 
as derivatives of the Green's function. 
By using Eq.~\eqref{eq:4.19} we can indeed rewrite the quantity in square brackets of Eq.~\eqref{eq:4.22} as
\begin{multline}
\int\,d3 G_{\kk}^{\(0\)}\(2,3\) \widetilde{\xi}_{\qq\gl}\(\rr_3\) G_{\kk-\qq}^{\(0\)}\(3,4\)= \\
 \overline{\partial_{\qq\gl}G_{\kk}\(2,4\)}.
\label{eq:4.23}
\end{multline}
We take now the static limit of $\chi^{\(0\)}$, which is consistent with the static and adiabatic approach taken in this section.  In this way, we obtain the contribution of the first diagram~(\ref{fig:12}.a) to $\gd\Sigma^{H}$
\begin{multline}
\gd\Sigma^{H,a}\(\rr_1\)= -\frac{1}{2} \sum_{\qq\gl} \int\,d\rr_2\,d\rr_3 v\(\rr_1,\rr_2\) \\
\times \partial_{\qq\gl}\chi^{\(0\)}\(\rr_2,\rr_3\) \widetilde{\xi}^{*}_{\qq\gl}\(\rr_3\)  \[2 n\(\go_{\qq\gl}\) +1\],
\label{eq:4.25}
\end{multline}
where the factor $1/2$ follows from the definition of $\chi^{\(0\)}$ (see Eq.~\eqref{eq:3.9}) that yields two equivalent contributions to $\gd\Sigma^{H,a}$.

If we now perform the same procedure to all bubble diagrams that contribute to time--dependent 
Hartree reducible response function we get the diagram~(\ref{fig:12}.b) and~(\ref{fig:12}.c). The final expression for $\gd\Sigma^{H}$ is
\begin{multline}
\gd\bgS^{H}\(\rr_1\)= \frac{1}{2} \sum_{\qq\gl} \int\,d\rr_2 \partial_{\qq\gl}\[\gee^{tdh}\(\rr_1,\rr_2\)\]^{-1}\\
\times \widetilde{\xi}_{\qq\gl}\(\rr_2\)  \[2 n\(\go_{\qq\gl}\) +1\].
\label{eq:4.26}
\end{multline}
We have demonstrated that, in the static and adiabatic limit, the series of diagrams due to the dressing of the electronic density as a consequence of the
electron--nucleus interaction reduce to a new contribution to the series of second--order diagrams. However, the derivative with respect to the atomic displacements is acting on the dielectric function.

The Eq.~\eqref{eq:4.26} allows us to rewrite the DFPT NRN potential $\widetilde{\gt}_{\qq\gl,\qq'\gl'}^{DFPT,NRN}$ in a purely MBPT language. 
A direct comparison between Eq.~\eqref{eq:4.26} and Eq.~\eqref{eq:4.15t} reveals that the difference between the MBPT and the DFPT approaches lies in the dielectric function ($\epsilon^{tdh}$ for the first and $\epsilon^{DFT}$ for the second).

%%%%%%%%%%%%%%%%%%%%%%%%%%%%%%%%%%%%%%%%%%%%%%%%%%%%%%%%%%%%%%%%%%%%%%%%%%%%%%%%%%%%%%%%%%%%%
\subsection{Screening of the first--order electron--nucleus interaction}
\label{sec:nth_order}
%%%%%%%%%%%%%%%%%%%%%%%%%%%%%%%%%%%%%%%%%%%%%%%%%%%%%%%%%%%%%%%%%%%%%%%%%%%%%%%%%%%%%%%%%%%%%
If we now consider terms beyond the Hartree approximation, the definition of the dielectric function that screens the second--order interaction will change. This means that if we add to the
Hartree potential a Fock operator, a $V_{Hxc}$ potential or a $GW$ self--energy, the dielectric function will
solve a time--dependent Fock, DFT or GW equations.

We can anyway add new correlation terms to the self--energy by keeping frozen the self--consistency at the DFT level. This means that the dielectric function will approximatively
solve a DFT equation that, within the limits of the discussion of Sec.~\ref{sec:screening}, will correspond to a simple RPA approximation written in the KS basis.

If we now include the Fan contribution to the total self--energy and 
follow the same procedure already discussed for the skeleton diagrams (see Sec.~\ref{sec:GW}), we will obtain that the $\xi$ function is
renormalized by the standard skeleton polarization diagrams. These, again, will be summed in a RPA 
dielectric function written in terms of KS bare Green's functions.

%%%%%%%%%%%%%%%%%%%%%%%%%%%%%%%%%%%%%%%%%%%%%%%%%%%%%%%%%%%%%%%%%%%%%%%%%%%%%%%%%%%%%%%%%%%%%
\subsection{Self--consistency }
%%%%%%%%%%%%%%%%%%%%%%%%%%%%%%%%%%%%%%%%%%%%%%%%%%%%%%%%%%%%%%%%%%%%%%%%%%%%%%%%%%%%%%%%%%%%%
As already mention earlier, the self--consistent diagrams play a crucial role in the definition of the second--order interaction. 

The screening of the rigid--nuclei second--order interaction can be obtained with a simple self--consistent Hartree approximation (that is 
already embodied in the DFT procedure). Instead, the non--rigid nuclei term requires to include, self--consistently, Fan diagrams and cannot be obtained with a simple
mean--field approximation.

This complicates the merging of MBPT and DFT as it is now clear that we can approximate the $\gt$ and $\xi$ 
functions with the DFPT counterparts under the condition that the Dyson equation is not solved self--consistently. Any kind of 
self--consistency would screen again the $\gt$ interaction leading to a severe over--screening.

The self--consistency must be taken from the solution of the KS equations and not re--introduced at the MB level.  This, as explained above, will correspond to the inclusion of certain 
class of diagrams, a practice that is well motivated within a many-body approach.

Finally, if the Dyson equation is not solved self--consistently, any additional electronic self--energy beyond the $V_{xc}$ can be introduced as an additive potential (see Eq.~\eqref{eq:1.9}). 
This means that it is safe to introduce non self--consistent quasi-particle corrections (electron and/or phonon mediated).

%%%%%%%%%%%%%%%%%%%%%%%%%%%%%%%%%%%%%%%%%%%%%%%%%%%%%%%%%%%%%%%%%%%%%%%%%%%%%%%%%%%%%%%%%%%%%
\subsection{Higher--order derivatives of the $\Vscf$}
%%%%%%%%%%%%%%%%%%%%%%%%%%%%%%%%%%%%%%%%%%%%%%%%%%%%%%%%%%%%%%%%%%%%%%%%%%%%%%%%%%%%%%%%%%%%%
The importance of creating a coherent merging of DFT and MBPT stems from the possibility of deriving an accurate and predictive approach to the \epip\, There is however an entire family of physical problems induced by a strong electron--phonon interaction that requires to introduce higher-order self--energy diagrams.
A crucial difference with respect to the state--of--the--art models (like Fr\"ohlich or Holstein Hamiltonians) is that the correct treatment of the 
nuclear positions produce a new category of diagrams of generic order $n$ where the n--th order derivative of the 
self--consistent potential appears
\begin{equation}
T^{\(n\)}_{R_1\cdots R_n}\(\rr\)\equiv \overline{\frac{\partial^n \Vscf\(\RR,\rr\)}{\partial{\RR_1}\cdots\partial{\RR_n}}},
\label{eq:5.7}
\end{equation}
where the Debye--Waller diagram is the lowest order example.
Thanks to the simplicity of DFPT, it is possible to define an iterative expression for the matrix elements of $T^{\(n\)}_{R_1\cdots R_n}\(\rr\)$
\begin{equation}
\left. T^{\(n\)}_{R_1\cdots R_n} \right|_{\nnpkkp}\equiv \la n\kk| T^{\(n\)}_{R_1\cdots R_n}\(\rr\)| n'\kk'\ra.
\label{eq:5.7p}
\end{equation}
Such matrix can be rewritten in terms of $T^{\(n-1\)}$ and of the n--th derivative of the KS energy levels
\begin{multline}
\left. T^{\(n\)}_{R_1\cdots R_n} \right|_{\nnpkkp}
= \sum\nolimits'_{m\pp, n\kk} \Biggl[
\frac{ \left.T^{\(1\)}_{R_1}\right|_{\nmkp}  \left. T^{\(n-1\)}_{R_2\cdots R_n}\right|_{\mnppkp}}{\gee_{m\pp}-\gee_{n\kk}}\\
- \frac{ \left.T^{\(n-1\)}_{R_2\cdots R_n}\right|_{\nmkp}  \left. T^{\(1\)}_{R_1}\right|_{\mnppkp}}{\gee_{n'\kk'}-\gee_{m\pp}}\Biggr]+
\PP_{1} \(\left. T^{\(n-1\)}_{R_2\cdots R_n}\right|_{\nnpkkp}\),
\label{eq:5.8}
\end{multline}
with 
\begin{equation}
\left. T^{\(0\)}\right|_{\mnppkp}=\gee_{n'\kk'}\gd_{n'm}\gd_{\kk' \pp},
\label{eq:5.9}
\end{equation}
and
\begin{equation}
\PP_{1} F\(\RR_1\dots\RR_n\)= -i \partial_{\RR_1}  F\(\RR_1\dots\RR_n\),
\label{eq:5.10}
\end{equation}
where $F$ is a generic function of the atomic positions.

More information on Eq.~\eqref{eq:5.8} is provided in appendix~\ref{appA}. Such procedure can be used to calculate numerically,
in an \ai\, manner, the high--order self--energies.

%%%%%%%%%%%%%%%%%%%%%%%%%%%%%%%%%%%%%%%%%%%%%%%%%%%%%%%%%%%%%%%%%%%%%%%%%%%%%%%%%%%%%%%%%%%%%
\section{Conclusions}
\label{sec:conclusions}
%%%%%%%%%%%%%%%%%%%%%%%%%%%%%%%%%%%%%%%%%%%%%%%%%%%%%%%%%%%%%%%%%%%%%%%%%%%%%%%%%%%%%%%%%%%%%
In this work we have studied the electron--phonon problem by comparing 
the standard Many--Body Perturbation Theory (MBPT) with Density Functional Theory (DFT) and Density Functional Perturbation Theory (DFPT).
By analyzing the different diagrams that contribute to the electronic self--energy we have achieved 
several important goals.

(i) The well--known electron--phonon induced tad--pole diagram is not zero in general but can be cancelled by a nuclear--nuclear self--energy if
the equilibrium nuclear positions are coherent with the level of correlation introduced at the single--particle levels. In the case of 
a non self--consistent calculation, the equilibrium positions evaluated with a DFT reference Hamiltonian removes the sum of the tad--pole diagrams.

(ii) Self--consistency diagrams dress the second--order interaction and the corresponding
Debye--Waller self--energy. This provides the many--body interpretation of the screened Debye--Waller self--energy already 
known in the DFPT case. 

(iii) We identify the specific series of diagrams that explain the non--rigid nuclei contribution to the Debye--Waller self--energy. 
The existence of this term was known only in a purely DFPT (static and adiabatic) approach.
In the present work, we provide a clear
physical interpretation of this term by performing a static limit of the MBPT expression to
demonstrate that, indeed, it reduces to the well--known DFPT result.

(iv) We have drawn a final series of statements regarding the possibility
to perform many--body perturbation theory calculations on top of  Density Functional and Density Functional Perturbation Theory avoiding 
the double counting of diagrams.

This work represents a firm and formally accurate inspection of the two methods while describing the limitations of the static DFPT approach and providing 
a practical way to go beyond by the merging with more advanced MBPT methods.

\section*{Acknowledgments}
%%%%%%%%%%%%%%%%%%%%%%%%%%%%%%%%%%%%%%%%%%%%%%%%%%%%%%%%%%%%%%%%%%%%%%%%%%%%%%%%%%%%%%%%%%%%%%%%%%%%%%%%%%%%%
Financial support for AM was provided by the {\em Futuro in Ricerca} grant No. RBFR12SW0J of the
Italian Ministry of Education, University and Research.
SP and XG gratefully acknowledge support from FRS-FNRS through a FRIA grant (SP), and through the PDR Grant T.0238.13 - AIXPHO .

\appendix
%%%%%%%%%%%%%%%%%%%%%%%%%%%%%%%%%%%%%%%%%%%%%%%%%%%%%%%%%%%%%%%%%%%%%%%%%%%%%%%%%%%%%%%%%%%%%%%%%%%%%%%%%%%%%%%%%%%%%%%%%%%%%%%%%%%%%%%%%%%%%%%%%%%%
\section{The interaction Hamiltonians in the phonon displacements representation}
\label{appB}
%%%%%%%%%%%%%%%%%%%%%%%%%%%%%%%%%%%%%%%%%%%%%%%%%%%%%%%%%%%%%%%%%%%%%%%%%%%%%%%%%%%%%%%%%%%%%%%%%%%%%%%%%%%%%%%%%%%%%%%%%%%%%%%%%%%%%%%%%%%%%%%%%%%%
We proceed here to discuss the details of the derivation and definition of the functions
\begin{equation}
\xi_{\gql}\(\rr\)= \partial_{\(\qq\gl\)} W_{e-n}\(\rr,\RR\),
\label{eq:B.1}
\end{equation}
\begin{equation}
\theta_{\gql,\qq'\gl'}\(\rr\)= \frac{1}{2}\partial^2_{\(\qq\gl\)\(\qq'\gl'\)} W_{e-n}\(\rr,\RR\),
\label{eq:B.2}
\end{equation}
\begin{equation}
\Xi_{\gql}= \partial_{\(\qq\gl\)} W_{n-n}\(\RR\),
\label{eq:B.3}
\end{equation}
\begin{multline}
\Theta_{\gql,\qq'\gl'}=\frac{1}{2} \partial^2_{\(\qq\gl\)\(\qq'\gl'\)} W_{n-n}\(\RR\) \\
- \Delta W_{n-n}^{ref}\(\RR\)\Big|_{\(\qq\gl\)\(\qq'\gl'\)},
\label{eq:B.4}
\end{multline}
where we purposely did not place a nucleus dependence $\RR$ on the left-hand side of those equations. We will indeed see that the dependence 
is lifted by Eq.~\eqref{eq:B.6}, as the derivatives have to be evaluated at $\overline{\RR}$.

Those functions enter in the different interaction terms of the \epi. We start by remembering that
\begin{equation}
W_{e-n}\(\rr,\RR\)=-\sum_{ls}Z_s v\(\rr-\RR_{ls}\),
\end{equation}
and
\begin{equation}
W_{n-n}\(\RR\)=\frac{1}{2}\sum_{ls,l's'}\nolimits'Z_s Z_{s'}v\(\RR_{ls}-\RR_{l's'}\).
\end{equation}
It follows that
\begin{equation}
\partial_{R_{ls\ga}} W_{e-n}\(\rr,\RR\)= Z_s v^{\(1\)}_{\ga}\( \rr-\RR_{ls} \),
\label{eq:B.5a}
\end{equation}
and
\begin{multline}
\partial^{2}_{R_{ls\ga}R_{l's'\ga'}} W_{e-n}\(\rr,\RR\)= \\
-Z_s v^{\(2\)}_{\ga\ga'}\( \rr-\RR_{ls} \)\gd_{ll'}\gd_{ss'},
\label{eq:B.5b}
\end{multline}
with $v^{\(1\)}_{\ga}\(\rr\)\equiv \partial_{r_{\ga}}v\(\rr\)$ and
$v^{\(2\)}_{\ga\ga'}\(\rr\)\equiv \partial^2_{r_{\ga}r_{\ga'}}v\(\rr\)$. By using Eqs.~\eqref{eq:B.5a} and \eqref{eq:B.5b} the
evaluation of the $\partial_{\(\qq\gl\)}$ and $\partial^2_{\(\qq\gl\)\(\qq'\gl'\)}$ is straightforward. It is enough to remember that
\begin{multline}
\partial_{\(\qq\gl\)} F\(\RR\)= 
\sum_{l s \ga} \(2 N M_s \go_{\qq \gl}\)^{-1/2}\\
\times \eta_{\ga}\(\qq\gl|s\) e^{i \qq\cdot\overline{\RR}_{ls}}
\partial_{R_{ls\ga}}\left.F\(\RR\)\right|_{\RR=\overline{\RR}},
\label{eq:B.6}
\end{multline}
with $F\(\RR\)$ a generic function. The resulting expression is therefore independent of $\RR$.
The $\partial^2_{\(\qq\gl\)\(\qq'\gl'\)}$ is then obtained by applying twice Eq.~\eqref{eq:B.6}. We finally obtain that
\begin{multline}
\xi_{\gql}\(\rr\)=\\
\sum_{ls\ga}\frac{Z_s e^{i \qq\cdot\oRR_{ls}}}{\sqrt{2N M_s\go_{\gql}}}  \eta_{\ga}\(\qq\gl|s\) v^{\(1\)}_{\ga}\( \rr-\oRR_{ls} \).
\label{eq:B.7}
\end{multline}
The same machinery can be applied to $\theta_{\gql,\qq'\gl'}\(\rr\)$ obtaining:
\begin{multline}
\theta_{\gql,\qq'\gl'}\(\rr\)=-\sum_{ls\ga\ga'}\frac{Z_s e^{i \(\qq+\qq'\)\cdot\oRR_{ls}}}{2N M_s \sqrt{\go_{\gql} \go_{\qq'\gl'} }}  \\
\times \eta_{\ga}\(\qq\gl|s\) 
\eta_{\ga'}\(\qq\gl|s\) 
v^{\(2\)}_{\ga\ga'}\( \rr-\oRR_{ls} \).
\label{eq:B.8}
\end{multline}
Two important quantities follow from the integral of Eq.~\eqref{eq:B.7} and \eqref{eq:B.8} when multiplied by two single--particle 
wavefunction: $\la n\kk |\xi_{\gql}\(\rr\) | n' \pp \ra$ and $\la n\kk |\gt_{\gql,\gqlp}\(\rr\) | n' \pp \ra$.
In order to evaluate those we have to consider, in the first-order derivative case, a term like
\begin{equation}
\sum_{ls}
\Biggl[\int d\rr\, \phi^{*}_{\nk}\(\rr\)v^{\(1\)}_{\ga}\( \rr-\oRR_{ls} \) \phi_{\npp}\(\rr\)
e^{i \qq\cdot\oRR_{ls}} \Biggr],
\label{eq:B.10}
\end{equation}
with the $\rr$ integral performed on the whole crystal. Now we observe that $v^{\(1\)}$ and $v^{\(2\)}$ depend on the atomic positions only via their argument.
If $\oRR_{ls}=\oRR_l+\tt_{s}$ with $\tt_s$ the position of the atom $s$ inside the unit cell located at $\oRR_l$
we can change variable from $\rr$ to $\rr'=\rr-\oRR_l$ to center the sum in the unit cell corner.  It follows that
Eq.~\eqref{eq:B.10} turns into
\begin{multline}
e^{i \qq\cdot\tt_{s}} \[\sum_{l}e^{i \(\qq+\pp-\kk\)\cdot\oRR_{l}}\]\\
\times \Biggl[\sum_{s\ga}\int_0 d\rr\, 
 u^{*}_{\nk}\(\rr\)v^{\(1\)}_{\ga}\( \rr-\tt_{s} \) u_{\npp}\(\rr\)e^{i \(\qq+\pp-\kk\)\cdot\rr}\Biggr],
\label{eq:B.11}
\end{multline}
where $u_{\nk}$ is the periodic part of the wavefunction.
By using the fact that $\sum_{l}e^{i \PP\cdot\oRR_l}=N\gd_{\PP}$ we finally obtain that
\begin{multline}
\la n\kk |\xi_{\gql}\(\rr\) | n' \pp \ra = \gd_{\pp,\kk-\qq} \la n\kk |\xi_{\gql}\(\rr\) | n' \kk-\qq\ra =\\
\sum_{s\ga}
\frac{Z_s e^{i \qq\cdot\tt_{s}}\sqrt{N}}{\sqrt{2 M_s\go_{\gql}}}  \eta_{\ga}\(\qq\gl|s\)\\
\times \[\int_0 d\rr\, u^{*}_{\nk}\(\rr\)v^{\(1\)}_{\ga}\( \rr-\tt_{s} \) u_{n'\kk-\qq}\(\rr\) \].
\label{eq:B.12}
\end{multline}
The same strategy can be used to define the matrix elements of $\gt$:
\begin{multline}
\la n\kk |\gt_{\gql,\gqlp}\(\rr\) | n' \pp \ra = \\
\gd_{\pp,\kk-\qq-\qq'} \la n\kk |\gt_{\gql,\gqlp}\(\rr\) | n' \kk-\qq-\qq'\ra =\\
-\sum_{s\ga\ga'}
\frac{Z_s e^{i \qq\cdot\tt_{s}}}{\sqrt{4 M_s\go_{\gql}}}  
\eta_{\ga}\(\qq\gl|s\)
\eta_{\ga'}\(\qq'\gl'|s\)\\
\times \[\int_0 d\rr\, u^{*}_{\nk}\(\rr\)v^{\(2\)}_{\ga\ga'}\( \rr-\tt_{s} \) u_{n'\kk-\qq-\qq'}\(\rr\) \].
\label{eq:B.13}
\end{multline}
The definition of $\Xi_{\gql}$ is similar and follows directly from Eq.~\eqref{eq:B.4} and from the extension of Eqs.~\eqref{eq:B.5a} and \eqref{eq:B.5b}
to the nucleus-nucleus potential:
\begin{multline}
\partial_{R_{ls\ga}} W_{n-n}\(\RR\)=\\
\sum_{l's'}\nolimits' Z_s Z_{s'}v_\alpha^{\(1\)}\(\RR_{ls}-\RR_{l's'}\),
\label{eq:B.14a}
\end{multline}
and
\begin{multline}
\partial^{2}_{R_{ls\ga}R_{l's'\ga'}} W_{n-n}\(\RR\)=\\
-Z_s Z_{s'}v^{\(2\)}_{\ga\ga'}\(\RR_{ls}-\RR_{l's'}\) (1-\gd_{ll'}\gd_{ss'})\\
+\sum_{l''s''}\nolimits' Z_s Z_{s''}v^{\(2\)}_{\ga\ga'}\(\RR_{ls}-\RR_{l''s''}\)\gd_{ll'}\gd_{ss'}.
\label{eq:B.14b}
\end{multline}
This, leads to 
\begin{multline}
\Xi_{\gql}\equiv
\sum_{ls,l's'}\nolimits' 
\frac{Z_{s}Z_{s'}}{\sqrt{2 M_{s} N \go_{\gql}}} 
e^{i\qq\cdot\oRR_{ls}} \\
\times  v^{\(1\)}_{\ga}\(\oRR_{ls}-\oRR_{l's'}\)
\eta_{\ga}\(\qq \gl|s\),
\label{eq:B.15}
\end{multline}
and, finally, that
\begin{widetext}
\begin{multline}
\frac{1}{2} \partial^2_{\(\qq\gl\)\(\qq'\gl'\)} W_{n-n}\(\oRR\)=
\sum_{ls\ga}\sum_{l's'\ga'} 
\frac{Z_s e^{i\(\qq\cdot\oRR_{ls}+\qq'\cdot\oRR_{l's'}\)}}{2N\sqrt{M_s M_{s'} \go_{\gql}\go_{\qq'\gl'}}} \eta_{\ga}\(\qq \gl|s\) \eta_{\ga'}\(\qq' \gl'|s'\)
\[ - Z_{s'} v^{\(2\)}_{\ga\ga'}\(\oRR_{ls}-\oRR_{l's'}\) (1-\gd_{ll'}\gd_{ss'}) \capo
+\sum_{l''s''}\nolimits' Z_{s''}v^{\(2\)}_{\ga\ga'}\(\oRR_{ls}-\oRR_{l''s''}\)\gd_{ll'}\gd_{ss'}\]
\label{eq:B.16}.
\end{multline}
\end{widetext}
%%%%%%%%%%%%%%%%%%%%%%%%%%%%%%%%%%%%%%%%%%%%%%%%%%%%%%%%%%%%%%%%%%%%%%%%%%%%%%%%%%%%%%%%%%%%%%%%
\section{An iterative expression for arbitrary n--th order derivatives of $V_{scf}$ with respect to the nuclear displacements}
\label{appA}
%%%%%%%%%%%%%%%%%%%%%%%%%%%%%%%%%%%%%%%%%%%%%%%%%%%%%%%%%%%%%%%%%%%%%%%%%%%%%%%%%%%%%%%%%%%%%%%%
One of the ingredients that are most difficult to calculate within an
{\em ab--initio} framework is the higher order derivatives of $\Vscf\(\RR,\rr\)$. While the first and second order are needed to calculate the lowest
order \ep self--energies an extension of the theory to the regime of strong interaction requires the knowledge of an arbitrary order derivative
\begin{equation}
T^{\(n\)}_{R_1\cdots R_n}\(\rr\)\equiv \overline{\frac{\partial^n \Vscf\(\RR,\rr\)}{\partial{\RR_1}\cdots\partial{\RR_n}}}.
\label{eq:A.1}
\end{equation}
In order to derive a close expression for $T^{\(n\)}$ we use the following property of the nuclear momentum operator $\PP_{I}$
\begin{equation}
\Bigl[ \PP_{1}, \Vscf\(\RR,\rr\)\Bigr]=-i \overline{\frac{\partial \Vscf\(\RR,\rr\)}{\partial{\RR_1}}}.
\label{eq:A.2}
\end{equation}
This identity can be iterated to give:
\begin{equation}
\Bigl[ \PP_{1},\Bigl[ \PP_{2}, \Vscf\(\RR,\rr\)\Bigr]\Bigr]= \overline{\frac{\partial^2
\Vscf\(\RR,\rr\)}{\partial{\RR_{1}\partial{\RR_{2}}}}}.
\label{eq:A.3}
\end{equation}
More generally we have that
\begin{equation}
T^{\(n\)}_{R_1 R_2\cdots R_n}=
i \Bigl[ \PP_{1},T^{\(n-1\)}_{R_2\cdots R_n} \Bigr]
\label{eq:A.4}
\end{equation}
In order to evaluate the $T^{\(n\)}$ matrix elements we start by noticing that, within DFPT,
\begin{equation}
\la n\kk|\PP_{1} |m\pp\ra = \(-i\)  \frac{\la n\kk|\frac{\partial \Vscf\(\RR,\rr\)}{\partial{\RR_{1}}}
|m\pp\ra}{\gee_{m\pp}-\gee_{n\kk}}.
\label{eq:A.5}
\end{equation}
Now we define
\begin{equation}
\left. T^{\(n\)}_{R_1\cdots R_n} \right|_{\nnpkkp}\equiv \la n\kk| T^{\(n\)}_{R_1\cdots R_n}\(\rr\)| n'\kk'\ra,
\label{eq:A.5p}
\end{equation}
and we plug into Eq.~\eqref{eq:A.4} a complete set of eigenstates obtaining
\begin{multline}
\left. T^{\(n\)}_{R_1\cdots R_n} \right|_{\nnpkkp}=\\
 i \sum_{m\pp\neq n\kk} \Biggl[ \brank \PP_{1} \( \ketmp \left. T^{\(n-1\)}_{R_2\cdots R_n}\right|_{\mnppkp} \)\\
 -\left.T^{\(n-1\)}_{R_2\cdots R_n}\right|_{\nmkp} \la m\pp|\PP_{1} |n'\kk'\ra
\Biggr].
\label{eq:A.6}
\end{multline}
In order to rewrite $T^{\(n\)}$ in terms of $T^{\(n-1\)}$ we need to evaluate 
\begin{multline}
\PP_{1} \( \ketmp \left. T^{\(n-1\)}_{R_2\cdots R_n}\right|_{\mnppkp} \)=\\
\Bigl(\PP_{1}  \ketmp\Bigr) \left. T^{\(n-1\)}_{R_2\cdots R_n}\right|_{\mnppkp}\\
 +\ketmp \( \PP_{1} \left. T^{\(n-1\)}_{R_2\cdots R_n}\right|_{\mnppkp} \).
\label{eq:A.7}
\end{multline}
Last step is, then, to use Eq.~\eqref{eq:A.5} to get
\begin{multline}
\left. T^{\(n\)}_{R_1\cdots R_n} \right|_{\nnpkkp}
= \sum_{m\pp \neq n\kk} \Biggl[
\frac{ \left.T^{\(1\)}_{R_1}\right|_{\nmkp}  \left. T^{\(n-1\)}_{R_2\cdots R_n}\right|_{\mnppkp}}{\gee_{m\pp}-\gee_{n\kk}}\\
- \frac{ \left.T^{\(n-1\)}_{R_2\cdots R_n}\right|_{\nmkp}  \left. T^{\(1\)}_{R_1}\right|_{\mnppkp}}{\gee_{n'\kk'}-\gee_{m\pp}}\Biggr]+
\PP_{1} \(\left. T^{\(n-1\)}_{R_2\cdots R_n}\right|_{\nnpkkp}\).
\label{eq:A.8}
\end{multline}
In order to check the soundness of this approach we apply it to the first order case:
\begin{multline}
\left. T^{\(1\)}_{R_1} \right|_{\nnpkkp}
= \sum_{m\pp \neq n\kk} 
\frac{ \left.T^{\(1\)}_{R_1}\right|_{\nmkp}  \left. T^{\(0\)}\right|_{\mnppkp}}{\gee_{m\pp}-\gee_{n\kk}}\\
- \sum_{m\pp\neq n\kk}
\frac{ \left.T^{0}\right|_{\nmkp}  \left. T^{\(1\)}_{R_1}\right|_{\mnppkp}}{\gee_{n'\kk'}-\gee_{m\pp}}+\PP_{1} \(\left. T^{\(0\)}\right|_{\nnpkkp}\).
\label{eq:A.9}
\end{multline}
But it is easy to verify that
\begin{equation}
\frac{\partial \Vscf\(\RR,\rr\)}{\partial{\RR_1}}=\frac{\partial \[T_{e}+\Vscf\(\RR,\rr\)\]}{\partial{\RR_1}},
\label{eq:A.10}
\end{equation}
which implies that
\begin{equation}
\left. T^{\(0\)}\right|_{\mnppkp}=\gee_{n'\kk'}\gd_{n'm}\gd_{\kk' \pp}.
\label{eq:A.11}
\end{equation}
Eq.~\eqref{eq:A.11},when  used in Eq.~\eqref{eq:A.9}, gives an identity 
that confirms the correctness of the iterative equation, Eq.~\eqref{eq:A.8}.

When Eq.~\eqref{eq:A.8} is applied to the second--order derivative it provides a well--known relation that connects the second to 
the first derivative~\cite{SP_2014,Gonze2011}.
\begin{multline}
\left. T^{\(2\)}_{R_1 R_2} \right|_{\nnkk}=\\
 \sum_{m\pp}{}^{'}\frac{ \left.T^{\(1\)}_{R_1}\right|_{\nmkp}  \left. T^{\(1\)}_{R_2}\right|_{\mnpk}+
\left.T^{\(1\)}_{R_2}\right|_{\nmkp}  \left. T^{\(1\)}_{R_1}\right|_{\mnpk}}
{\gee_{m\pp}-\gee_{n\kk}}\\
-\partial^2_{\RR_1\RR_2}\gee_{n\kk}\(\RR\).
\label{eq:A.13}
\end{multline}
More in general Eq.~\eqref{eq:A.8} can efficiently rewrite the n--th order derivative in terms of the first order derivative and of the n--th order derivative of the
electronic energies, 
$\frac{\partial^n\gee_{n\kk}\(\RR\)} {\partial\RR_1\cdots\partial\RR_n}$. 
But these derivatives can be efficiently calculated using DFPT or finite differences~\cite{SP_2014,Antonius2014}.

\bibliographystyle{apsrev4-1}
\bibliography{paper}

\end{document}